\documentclass[superscriptaddress,twocolumn,showpacs,prb]{revtex4}

\usepackage{graphicx,amsmath}
\usepackage{dcolumn}
\usepackage{amssymb}
\usepackage{bm}
\usepackage{color}
\newcommand{\av}{_\text{av}}
\newcommand{\bfk}{\boldsymbol{k}}
\newcommand{\crit}{_\text{c}}

\begin{document}

\title{
The correspondence between long-range and short-range spin glasses.
}

\author{R.~A.~Ba\~nos}
\affiliation{Departamento
de F\'\i{}sica Te\'orica, Universidad de
Zaragoza, 50009 Zaragoza, Spain.}
\affiliation{Instituto de Biocomputaci\'on y
F\'{\i}sica de Sistemas Complejos (BIFI), Zaragoza, Spain.}

\author{L.~A.~Fernandez}
\affiliation{Departamento
de F\'\i{}sica Te\'orica I, Universidad
Complutense, 28040 Madrid, Spain.}
\affiliation{Instituto de Biocomputaci\'on y
F\'{\i}sica de Sistemas Complejos (BIFI), Zaragoza, Spain.}

\author{V.~Martin-Mayor}
\affiliation{Departamento de F\'\i{}sica
Te\'orica I, Universidad Complutense, 28040 Madrid, Spain.}
\affiliation{Instituto de Biocomputaci\'on y 
F\'{\i}sica de Sistemas Complejos (BIFI), Zaragoza, Spain.}

\author{A.~P.~Young}
\affiliation{Department of Physics, University of California,
Santa Cruz, California 95064}

\date{\today}

\begin{abstract}
We compare the critical behavior of the short-range Ising spin glass with a spin
glass with long-range interactions which fall off as a power $\sigma$ of the
distance. We show that there is a value of $\sigma$ of the long-range model
for which the critical
behavior is very similar to that of the short-range model in four dimensions.
We also study a value of $\sigma$ for which we 
find the critical behavior to be compatible with that of the
three dimensional model, though we have much less precision
than in the four-dimensional case.
\end{abstract}
\maketitle

\section{Introduction}
\label{sec:intro}

In the theory of systems at their critical point it is instructive
to consider a range of dimensions $d$, since above an upper critical
dimension, $d_u$, the critical behavior becomes quite simple and
corresponds to that
of mean field theory. Hence it is desirable to understand critical behavior up
to, and just above, $d = d_u$. For the case of spin glasses,\cite{binder:86}
where much of what we
know has come from numerical simulations, this has been difficult because (i)
the value of $d_u$ is quite large ($d_u = 6$ as opposed to 4 for conventional
systems like ferromagnets) and (ii) slow dynamics, coming from the complicated
``energy landscape", prevents equilibration of systems with more than of
order $10^4$ spins at and below the
transition temperature $T\crit$. Since the total number of spins $V$ is related
to the linear size $L$ by $V = L^d$, for dimensions around $d_u\, (= 6)$ it is
then not possible to study a \textit{range} of values of $L$, which, however, is
necessary to carry out a finite-size scaling\cite{privman:90,amit:05} (FSS)
analysis. 

It has been proposed\cite{katzgraber:03} to try to circumvent this problem by using,
instead, a one-dimensional spin glass model in which the interactions $J_{ij}$
fall off as a power of the distance, roughly $J_{ij} \sim 1/|r_i - r_j|^\sigma$, since
varying $\sigma$ in this 1-$d$ model seems to be analogous to varying $d$ in a
short-range models. In both cases there is a range where there is no
transition ($d$ less than a lower critical dimension $d_l$, $\sigma$ greater
than a certain value $\sigma_l$), a range where there is a transition with
non-mean field exponents ($d_l < d < d_u$, $\sigma_l > \sigma > \sigma_u$ for
a certain $\sigma_u$ which turns out to be 2/3), and a transition with mean
field exponents ($d_u < d < \infty$, $\sigma_u > \sigma > 1/2$). The
advantage of the 1-$d$ model
is that one can study a large range of linear sizes for the whole
range of $\sigma$. Consequently, there have been several subsequent
studies\cite{katzgraber:03b,katzgraber:05,katzgraber:09,larson:10,sharma:11a,leuzzi:08,leuzzi:09,leuzzi:11}
on these models. 

The question that we tackle here is whether this connection between long-range
models in 1-dimension and short range models in a range of dimensions is just a vague
analogy or whether the connection can be made precise in the following sense:
for a given $d$ is there a value of $\sigma$ such that \textit{all} the
critical exponents of the short-range model correspond with those of the
long-range model (in the sense of Eq.~\eqref{SR_LR} below)? We will denote the
value of $\sigma$ in Eq.~\eqref{SR_LR} as a \textit{proxy} for the dimension
$d$.

A relation between the long-range (LR) and short-range (SR) exponents has been
proposed in Ref.~\onlinecite{larson:10}. We reproduce their argument here in a
more general formulation. Consider the singular part of the free energy
density. For a system in $d$ dimensions it has the scaling form
\begin{equation}
f_\text{sing} = \frac{1}{L^d} \widetilde{f}\left(L^{y_T} t, L^{y_H} h, L^{y_u}
u \right) \, ,
\label{fsing}
\end{equation}
where $\widetilde{f}$ is a scaling function, $t \equiv (T - T\crit)/T\crit$ is the
reduced temperature, $h$ is the magnetic field (for a spin glass it is
actually the variance of a random field), $u$ is the operator
which gives the leading correction to scaling, $y_T$ is the thermal exponent,
$y_H$ is the magnetic exponent, and $y_u \ (< 0)$ is the exponent for the
leading correction to scaling. These exponents can be expressed in
terms of more commonly used exponents, 
\begin{equation}
y_T = \frac{1}{\nu}, \qquad y_H = \frac{1}{2}\, (d +2 - \eta), \qquad
y_u = -\omega \, ,
\end{equation}
where $\nu$ is the correlation length exponent, $\eta$ describes the power-law
decay of correlations at the critical point, and $\omega > 0$.

We make a connection between the two models by equating the singular part
of their free energy densities, i.e.
\begin{multline}
\frac{1}{L^d} \widetilde{f}_\mathrm{SR}\left(L^{y_T^\mathrm{SR}} t, L^{y_H^\mathrm{SR}} h,
L^{y_u^\mathrm{SR}} u \right)  =  \\
\frac{1}{L} \widetilde{f}_\mathrm{LR}\left(L^{y_T^\mathrm{LR}} t, L^{y_H^\mathrm{LR}} h,
L^{y_u^\mathrm{LR}} u \right).
\end{multline}
In order to compare exponents we need to eliminate the 
different prefactors in front of
the scaling functions by writing everything in terms of the total number of
spins $V$ where $V = L^d$ for SR and $V = L$ for LR. Canceling a factor of
$1/V$ on both sides gives
\begin{multline}
\widetilde{f}_\mathrm{SR}\left(V^{y_T^\mathrm{SR}/d} t, V^{y_H^\mathrm{SR}/d} h,
V^{y_u^\mathrm{SR}/d} u \right)  = \\
\widetilde{f}_\mathrm{LR}\left(V^{y_T^\mathrm{LR}} t, V^{y_H^\mathrm{LR}} h,
V^{y_u^\mathrm{LR}} u \right).
\end{multline}
Hence, for each of the exponents,
the correspondence between the LR and SR values is
\begin{equation}
y_\mathrm{LR}(\sigma) = \frac{y_\mathrm{SR}(d)}{d} \, .
\label{SR_LR}
\end{equation}
We note that in the mean-field regime, $6 < d < \infty, 2/3 > \sigma > 1/2$,
Eq.~\eqref{SR_LR} holds consistently\cite{larson:10}
for the thermal, magnetic, and correction
exponents with
\begin{equation}
d = \frac{2}{2 \sigma - 1}, \qquad \text{(mean\ field\ regime)}\, ,
\end{equation}
since\cite{harris:76,kotliar:83} $\eta_\mathrm{SR} = 0, \eta_\mathrm{LR} = 3 -
2 \sigma, \nu_\mathrm{SR} = 1/2, \nu_\mathrm{LR} = 1/(2\sigma - 1),
\omega_\mathrm{SR} = (d - 6)/2,$ and $\omega_\mathrm{LR} = 2 - 3
\sigma$. Furthermore, the exponents also match to first order in $6-d$ for the
SR model and $\sigma - 2/3$ for the LR model.\cite{moore:pc} Actually,
Eq.~\eqref{SR_LR} (at least as applied to the thermal exponent $\nu=1/y_T$) can
be derived for all $d$ and $\sigma$ from a super-universality
hypothesis.\cite{parisi:pc}

In this paper we will investigate whether, for $d = 3$ and 4, we can
find a value of $\sigma$ which satisfies Eq.~\eqref{SR_LR} simultaneously for the thermal,
magnetic and correction to scaling exponents.

One advantage of long-range systems is that the exponent $\eta$ is known
exactly as was first shown by Fisher et al.\cite{fisher:72} for ferromagnets.
The result for spin glasses is 
\begin{equation}
2 - \eta_\mathrm{LR}(\sigma) = 2 \sigma - 1\, ,
\label{eta_LR}
\end{equation}
so Eq.~\eqref{SR_LR} for the magnetic exponent $y_H\, (= (d+2-\eta)/2)$
can be written\cite{larson:10}
\begin{equation}
2 \sigma - 1 = \frac{2 - \eta_\mathrm{SR}(d)}{d} \, ,
\label{sigma_eta}
\end{equation}
which immediately gives us a value of $\sigma$ which acts as a proxy for $d$
provided we know $\eta_\mathrm{SR}(d)$. For $d = 3$, the value of $\eta_\mathrm{SR}$, as well as other
exponents, has been
determined accurately by Hasenbusch et al.\cite{hasenbusch:08b} and we use
their values here. In particular, they find $\eta_\mathrm{SR}(3) = -0.375(10)$, which,
according to Eq.~\eqref{sigma_eta}, corresponds to a proxy value
$\sigma = 0.896$. We shall
therefore perform simulations for this value of $\sigma$ to see if the other
exponents, $y_T$ and $y_u$, also match those of the $d=3$
results\cite{hasenbusch:08b}
according to Eq.~\eqref{SR_LR}.

However, for $d = 4$, the values of $\eta_\mathrm{SR}$ and the
other exponents are not known with great precision, so we carry here out a
careful study of this model here to determine them more accurately.  We 
find $\eta_\mathrm{SR}(4) = -0.320(13)$ for which the proxy value of $\sigma$,
according to Eq.~\eqref{sigma_eta}, is $\sigma = 0.790$. We therefore also
study this value of $\sigma$ so see if the other exponents match those of the
$d=4$ simulations according to Eq.~\eqref{SR_LR}.

It is also convenient to note that Eq.~\eqref{SR_LR} for the thermal
exponent $y_T\ (= 1 / \nu)$ can be written
\begin{equation}
\nu_\mathrm{LR}(\sigma) = d \, \nu_\mathrm{SR}(d) \, ,
\label{nu_compare}
\end{equation}
and, since $y_u = -\omega$, the connection between the correction to scaling
exponents is
\begin{equation}
\omega_\mathrm{LR}(\sigma) = \frac{\omega_\mathrm{SR}(d)}{d}\, .
\label{omega_compare}
\end{equation}
To summarize, the main goal of this paper is to see if there
is a single value of $\sigma$ which simultaneously satisfies
Eqs.~\eqref{sigma_eta}, \eqref{nu_compare}, and \eqref{omega_compare} for
$d=3$ and (with a different value of $\sigma$) for $d=4$.

The plan of this paper is as follows. In Sec.~\ref{sec:model} we describe the model
and the observables we calculate. Section \ref{sec:fss} discusses the
finite-size scaling analysis, while Sec.~\ref{sec:simulation} describes the
details of the simulations. The results and analysis are presented in
Sec.~\ref{sec:results}, while our conclusions are summarized in
Sec.~\ref{sec:conclusions}.

\section{Model and Observables}
\label{sec:model}

We consider the Edwards-Anderson spin-glass model with Hamiltonian
\begin{equation}
\mathcal{H} = - \sum_{\langle i, j \rangle} 
J_{ij} S_i S_j \, ,
\label{ham}
\end{equation}
where the Ising spins $S_i$ take values $\pm 1$ and the quenched interactions
$J_{ij}$ are
independent random variables, the form of which will be different for the different
models that we study.

The first model is a nearest neighbor spin glass in four dimensions in which the
$J_{ij}$ take values $\pm 1$ with equal probability if $i$ and $j$ are nearest
neighbors, and are 0 otherwise, i.e.\ the probability distribution is
\begin{equation}
P(J_{ij}) = \left\{
\begin{array}{ll}
\frac{1}{2} \left[ \delta(J_{ij} - 1) + \delta(J_{ij} + 1)
\right], & \text{($i,j$ neighbors)}, \vspace{3mm}\\
\delta(J_{ij}), & \text{(otherwise)}.
\end{array}
\right.
\end{equation}
The advantage of the $\pm 1$ interactions is that we are able to use multispin
coding,\cite{newman:99} in which the interactions and the spins are represented by a single
bit rather than a whole word. In fact, our C code uses 128-bit words, using
the streaming SIMD extensions,
so we simulate 128 samples in parallel. In order to
gain the full speedup, we use the \textit{same} random numbers for each of the
128 samples in a ``batch''. Hence, while
the results for each sample are unbiased, there may be correlations between
samples in the same batch. Consequently,
when we estimate error bars we first average over the samples in a batch and
use this average as a single data point in the analysis. Data from different
batches are uncorrelated.

The spins are on a 4-dimensional hypercubic lattice of linear size $L$ 
with periodic boundary
conditions. The total number of spins is $V = L^4$.

The description of the interactions we take for the 1-$d$ models is a bit more
complicated.
The interactions must fall off with distance such that
\begin{equation}
[ J_{ij}^2 ]\av \propto \frac{1}{r_{ij}^{2\sigma} }\, ,
\label{lrints}
\end{equation}
where $r_{ij} = | r_i - r_j |$ (the $i=0,1,\ldots L-1$ sites in the graph are
placed in a circle of radius $L/(2\pi)$, site $i$ is at angle $i 2\pi/L$). On the other hand
$[ \cdots ]\av$ denotes an average over the interactions.  The simplest way
to do this is to have every spin interact with every other spin with an
interaction strength which has zero mean and standard deviation $\propto
1/r_{ij}^\sigma$. However, this is inefficient to simulate for large sizes,
because the CPU time per sweep is of order $L^2$, rather than $L z$ in
short-range systems with coordination number $z$.  Fortunately, it was
realized by Leuzzi et al.,\cite{leuzzi:08} that one can have the CPU time
scale also like $L z$ for the long range model if one dilutes it. In their
version, most interactions are zero and those that are non-zero have a
strength of unity (i.e.\ the strength does not decrease with distance). Rather
it is the \textit{probability} of the interaction being non-zero which
deceases with distance. In the specific construction of Leuzzi et
al.\cite{leuzzi:08} there are a total of $L z / 2$ non-zero interactions with
an \textit{average} degree (i.e.\ coordination number) of $z$ and the
probability of a non-zero interaction given by
\begin{equation}
p_{ij} = 1 - \exp(-A / r_{ij}^{2\sigma}) 
\ (\simeq A / r_{ij}^{2\sigma}\ \text{at large $r_{ij}$}) \, ,
\label{pij}
\end{equation}
where $A$ is chosen so that the mean degree is equal to some specified value $z$.

In the Leuzzi et al model, the
degree is not the same for all sites but has a Poisson distribution with
mean $z$. Since we wish to implement multispin coding, and since the computer
code for this depends strongly on the degree (and gets complicated for large
degree), we study, instead, a model with \textit{fixed} degree.

\begin{figure}
\includegraphics[width=\columnwidth]{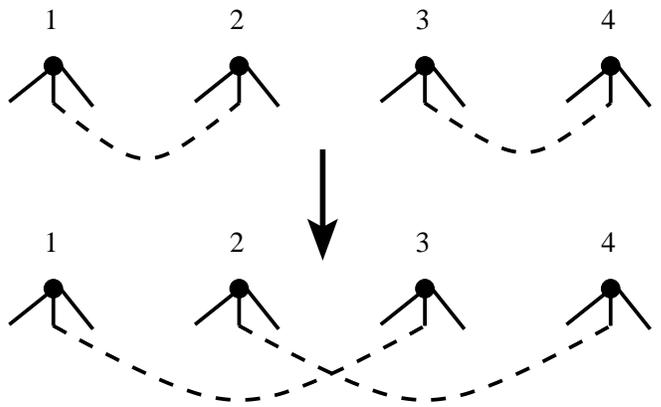}
\caption{Each
site has a fixed number of ``legs'' (here we show three) and these legs
are paired up by ``edges''. In the top row, one edge connects sites 1 and 2,
and another edge connects sites 3 and 4. A basic Monte Carlo move for the
bond-generation simulation consists of
reconnecting two edges, as shown in the bottom row.  (Other edges are present
but not shown.)
}
\label{fig:join}
\end{figure}

We are not aware of any simple algorithm to generate bonds of arbitrary length
such that each site has a specified number of bonds ($z$ here) and the
probability of a bond between $i$ and $j$ varies with distance $r_{ij}$ in
some specified way ($\propto 1/r_{ij}^{2\sigma}$ here). We therefore construct
the Hamiltonian for which we will simulate the spins by \textit{first}
performing a Monte Carlo simulation \textit{of the bonds}. A similar (but
simpler) problem was resolved in this way in
Ref.~\onlinecite{fernandez:10}. We take the ``Hamiltonian'' of the bonds to be
given by
\begin{equation}
\mathrm{e}^{-\mathcal{H}_\text{bond}}=
\mathrm{e}^{ -\sum_{\langle i,j \rangle}
\epsilon_{ij} \log r_{ij}^{2\sigma}}\prod_k\delta\big(\sum_l \epsilon_{kl}-z\big),
\label{Hbond}
\end{equation}
where $\epsilon_{ij} = 0$ or $1$, in which $1$ represents
a bond present between sites
$i$ and $j$, and $0$ represents no bond. Graphically, we regard 
each site $i$ as having $z$
``legs'' associated with it, and we initially pair up the legs in
a random way, representing each connected pair graphically as an ``edge''
and giving the value $\epsilon_{ij} = 1$ to all edges while all other pairs
$(i,j)$ have $\epsilon_{ij} = 0$.
We then run a Monte Carlo simulation in which the non-zero
$\epsilon_{ij}$ are swapped according to a Metropolis probability for the
Hamiltonian in Eq.~\eqref{Hbond}. To maintain exactly $z$ non-zero $\epsilon$'s
for each site the basic move involves reconnecting \textit{two} bonds as shown in the
sketch in Fig.~\ref{fig:join}.

Specifically, we first choose site 1 in Fig. 1, with uniform probability among
the $L$ possible choices. Next, site 3 is chosen with probability proportional
to $1/r^{2\sigma}$ ($r$ is the distance among sites 1 and 3).  Finally, site 2
(site 4) is chosen with uniform probability among the $z$ "neighbors" of site
1 (site 3). Before the move is attempted, we need to check that the sites
1, 2, 3 and 4 verify two consistency conditions. First, the four sites should be
all different. Second, we require that neither sites 1 and 4, nor 2 and 3, are
paired. If the consistency conditions are met, the basic move can be attempted
and then be accepted/rejected with Metropolis probability. One
\emph{sweep} corresponds to
$L z$ selection of sites of type "1" in Fig.~\ref{fig:join}.

After a suitable equilibration time,
\footnote{We run the Monte Carlo of the bonds for one million sweeps. We are
confident about graph-equilibration because we compared the outcome of
widely differing starting points for the simulation: either a graph with the
topology of a crystal with periodic boundary conditions, or the random graph
described in the main text. For either type of starting point, we compared
several graph-properties, in particular 
the bond-length distribution and the
``Hamiltonian" defined in Eq. (15). In all cases studied, we found
that memory of the starting configuration was lost after $10^5$ sweeps,
but simulated for a total $10^6$ sweeps to be on the safe side.}
we freeze the $\epsilon_{ij}$, and the resulting
set of non-zero $\epsilon_{ij}$ defines a ``graph''. Each of the 128 samples
in a single batch of the multispin coding
algorithm has the \textit{same} graph. On the edges of
the graph we put interactions with values $\pm 1$ with equal probability
chosen \textit{independently} for each edge \textit{in
each sample} in a batch. The result is that the probability
distribution \textit{for a single bond} is given by
\begin{equation}
P(J_{ij}) = (1 - p_{ij})\, \delta(J_{ij}) + p_{ij} \, \frac{1}{2} \,
\left[ \delta(J_{ij} - 1) + \delta(J_{ij} + 1)\right]\, ,
\end{equation}
in which $p_{ij}$ is given \textit{approximately} by Eq.~\eqref{pij} for an
appropriate choice of $A$ corresponding to the specified value of $z$. However, the
bonds are no longer statistically independent; rather there are
\textit{correlations} which ensure that each site has \textit{exactly}
$z$ non-zero bonds. 
For both $\sigma = 0.896$ and $0.790$ we take $z=6$ neighbors.

We now describe the quantities that we calculate in the simulations.
The spin glass order parameter is 
\begin{equation}
q = \frac{1}{V} \sum_{i=1}^V S_i^{(1)} S_i^{(2)} \, ,
\end{equation}
where ``(1)" and ``(2)'' are two identical copies of the system with the same interactions.
Its Fourier transform to wavevector $\bfk$ is denote by $q(\bfk)$.  We will
calculate the spin glass susceptibility
\begin{equation}
\chi_\text{SG} = V [\, \langle q^2 \rangle\, ]\av \, ,
\label{chiSG}
\end{equation}
and also its wavevector-dependent generalization,
\begin{equation}
\chi_\text{SG}(\bfk) = V [\, \langle |q(\bfk)|^2 \rangle\, ]\av \, .
\end{equation}
From this we can extract the correlation
length,\cite{cooper:82,palassini:99b,ballesteros:00,amit:05}
\begin{equation}
\xi_L = \frac{1}{2 \sin(\pi/L)} \,\,
\sqrt{\frac{\chi(0)}{\chi({\bfk}_1)}-1} \, ,
\label{xiL}
\end{equation}
where $\bfk_1$ is the smallest non-zero wavevector, $\bfk_1 = (2\pi/L)(1,0,0,0)$ for the 4-$d$ model
and $k_1 = 2 \pi/L$ for the long-range models in 1-$d$.
Other quantities that we calculate, important because
they are dimensionless like $\xi_L/L$, are
the moment ratios,
\begin{equation}
\label{U4}
U_4 = \frac{ [\, \langle q^4 \rangle\, ]\av }{ [\, \langle q^2 \rangle\, ]\av^2 }
\, , 
\end{equation}
\begin{equation}
U_{22} = \frac{ [\, \langle q^2 \rangle^2\, ]\av
- [\, \langle q^2 \rangle\, ]\av^2 
}{ [\, \langle q^2 \rangle\, ]\av^2 }
\, , 
\label{U22}
\end{equation}
and the susceptibility ratio
\begin{equation}
R_{12} = \frac{\chi_{SG} (\bfk_1)}{\chi_{SG}(\bfk_2)} \, ,
\label{R12}
\end{equation}
where $\bfk_2$ is the second smallest non-zero wavevector, $\bfk_2 =
(2\pi/L)(1,1,0,0)$ for the 4-$d$ model 
and $k_2 = 4 \pi/L$ for the long-range models.
We will also determine derivatives with respect to $\beta$ of several of these
quantities using the result 
\begin{equation}
\left\langle\, \frac{\partial O}{\partial \beta}\, \right\rangle = \langle O \, \mathcal{H}
\rangle - \langle O \rangle \, \langle \mathcal{H} \rangle \, .
\label{beta_deriv}
\end{equation}

\section{Finite-Size Scaling Analysis}
\label{sec:fss}

Using data from finite-sizes, we have to extract the transition temperature
$T\crit$, the correction to scaling exponent $\omega$ (since corrections to
scaling are significant), the correlation length exponent $\nu$, and (for the
short-range model which it's value is not known exactly) the exponent $\eta$.
In this section we show how to include the \textit{leading} correction to
FSS. There are several sources of subleading corrections which will not be
included in the formulae in this section, though we will try to include them
empirically in some of the fits to the data, as discussed later in the
section. 

It is desirable to compute the various quantities 
\textit{one at a time} so the value of the
exponents depend on each other to the least extent possible.
We therefore adopt the following
procedure.

We start with the finite-size scaling (FSS) form of a \textit{dimensionless}
quantity, since these  quantities are simpler to analyze than those with dimensions
and so they form the core of our analysis.

Dimensionless quantities are scale-invariant, which means that at $T\crit$ they remain finite
(neither zero nor infinite) in the limit of large $L$. However dimensionless
quantities are not only
scale-invariant, they are also \textit{universal} (i.e. they remain constant under
Renormalization-Group transformations). Examples of dimensionless quantities
are $\xi_L/L$, $U_4$, $U_{22}$ and $R_{12}$. The distinction among
scale-invariant and dimensionless quantities has been stressed in
Ref.~\onlinecite{hasenbusch:08b}.  Here we will discuss dimensionless quantities, but will
comment on quantities which are scale-invariant but not dimensionless
in the last paragraph of this section.

A dimensionless
quantity $f(L, t)$ has the FSS scaling form\cite{binder:81b,ballesteros:96a,ballesteros:00}
\begin{equation}
f(L, t) = \widetilde{F}_0(L^{1/\nu} t) + L^{-\omega} \widetilde{F}_1(L^{1/\nu} t),
\label{FSS_nodimen}
\end{equation}
where $\omega$ is the correction to scaling exponent, and 
\begin{equation}
t = {T - T\crit \over T\crit}.
\label{t}
\end{equation}
We
are interested in the behavior at large $L$ and small $t$, and including just
the leading corrections in $1/L$ and $t$ gives
\begin{equation}
f(L, t) \simeq  \widetilde{F}_0(0)  + L^{1/\nu} t\, \widetilde{F}'_0(0)
+ L^{-\omega} \widetilde{F}_1(0)   .
\label{fscale}
\end{equation}
It will be useful to determine the values of $t^*_L$ where 
the quantity $f$ takes the same value for sizes $L$ and $s L$, where
$s$ is a scale factor which we shall take to be 2 here. We have
\begin{multline} 
\widetilde{F}_0(0)  + L^{1/\nu} t^\star_L\, \widetilde{F}'_0(0)
+ L^{-\omega} \widetilde{F}_1(0) = \\
\widetilde{F}_0(0)  + (s L)^{1/\nu} t^\star_L\,
\widetilde{F}'_0(0) + (s L)^{-\omega} \widetilde{F}_1(0) ,
\end{multline}
which gives
\begin{equation}
{T^\star_L - T\crit \over T\crit} \equiv t^\star_L = A_s^f L^{-\omega - 1/\nu} \, ,
\end{equation}
or equivalently, to leading order,
\begin{equation}
{\beta\crit - \beta^\star_L \over \beta\crit} = A_s^f L^{-\omega - 1/\nu} \, ,
\label{t*}
\end{equation}
where the non-universal amplitude is given by
\begin{equation}
A_s^f = \frac{(1 - s^{-\omega}) \, F_1(0) }{ (s^{1/\nu} - 1) \, F'_0(0) } \, .
\label{Asf}
\end{equation}
One can use Eq.~\eqref{t*} to locate $\beta\crit$. As we shall see, the exponents
$\omega$ and $1/\nu$ are determined separately, and we use those values when
fitting the data to Eq.~\eqref{t*}.

We shall determine the critical exponents using the
quotient method,\cite{ballesteros:96a} which is a more modern form
of Nightingale's phenomenological renormalization.\cite{nightingale:76} First
we determine the correction exponent $\omega$ by
applying the quotient method to
dimensionless quantities. Consider a second dimensionless quantity $g(L, t)$
which varies near $T\crit$ in the same way as $f$ in Eq.~\eqref{fscale}, i.e.
\begin{equation}
g(L, t) \simeq  \widetilde{G}_0(0)  + L^{1/\nu} t\, \widetilde{G}'_0(0)
+ L^{-\omega} \widetilde{G}_1(0)   .
\label{gscale}
\end{equation}
Now compute $g(L, t)$ at $t^\star_L$, given by Eq.~\eqref{t*},
the temperature where results for $L$
and $s L$ intersect for some \textit{different} dimensionless quantity $f$. We have
\begin{equation}
g(L, t^\star_L) \simeq  \widetilde{G}_0(0) 
+ A_s^{g,f} L^{-\omega} ,
\end{equation}
where $A_s^{g,f} = A_s^f \widetilde{G}'_0(0) + \widetilde{G}_1(0)$.
While this could be used directly to determine $\omega$ it is more convenient
to take the ratio (quotient) of this result with the corresponding result for
size $s L$, i.e.
\begin{equation}
Q(g) \equiv 
\frac{g(s L, t^\star_L) }{g(L, t^\star_L)} = 1 + B_s^{g, f} L^{-\omega} \, ,
\label{quotient_omega}
\end{equation}
where the amplitude $B_s^{g, f}$ is non-universal (because of the definition, it 
is zero if the quantities $f$
and $g$ are the same). Eq.~\eqref{quotient_omega} 
is the most convenient expression from which to determine $\omega$ since it
just involves the one unknown exponent $\omega$, and one amplitude $B$. These
quantities can be
determined by a straight-line fit to a log-log plot of $Q(g) - 1$ against $L$.

To determine the other exponents $\nu$ and $\eta$ we need to consider the FSS
scaling form of quantities which \textit{have} dimensions. Consider some
quantity $O$ which diverges in the bulk like $t^{-x_O}$. Including the leading
correction it has the FSS form
\begin{equation}
O(L, t) = L^{y_O} \left[ \widetilde{O}_0(L^{1/\nu} t) + L^{-\omega} \widetilde{O}_1(L^{1/\nu} t)
\right] \, ,\label{FSSleading}
\end{equation}
where $y_O = x_O /  \nu$. Repeating the above arguments, and determining $O$
for sizes $L$ and $s L$ at the intersection temperature $t^\star_L$ for the dimensionless
quantity $f$ for sizes $L$ and $s L$, the quotient can be written as
\begin{equation}
Q(O) \equiv
\frac{O\left (sL,t^\star_L\right)}
{O\left (L,t^\star_L\right)}=s^{y_O} + B_s^{O, f}\, L^{-\omega}\, .
\label{QUOTIENTS}
\end{equation}
Using the value of
$\omega$ determined from Eq.~\eqref{quotient_omega} the exponent $y_O$ is
determined from Eq.~\eqref{QUOTIENTS} by a straight line fit to a plot of 
$Q(O)$ against $1/L^\omega$. 

To determine $\eta$ we can use Eq.~\eqref{QUOTIENTS} for the spin-glass
susceptibility $\chi_{SG}$, since $y_O = 2 - \eta$ because the susceptibility
exponent $\gamma\ (\equiv x_{\chi_{SG}}) = (2-\eta)\nu$.  To determine $\nu$
we note that $\xi_L/L$ is dimensionless and so has the same FSS scaling form
as in Eq.~\eqref{FSS_nodimen}.  Differentiating, for instance, $\xi_L$ with respect to
$\beta$ brings down a factor of $L^{1/\nu}$ and so $y_O = 1 + 1/\nu$ in this
case ($y_O = 1/\nu$ if we take the \textit{logarithmic} derivative). Hence we determine $1 +
1/\nu$ from Eq.~\eqref{QUOTIENTS} with $O$ given by the $\beta$ derivative of
$\xi_L$.

To conclude, to carry out the FSS analysis we do the following steps:
\begin{enumerate}
\item
\label{stage1}
Determine $\omega$ from Eq.~\eqref{quotient_omega} for one or more
dimensionless quantities $f$.
\item
\label{stage2}
Using the value of $\omega$ so determined,
obtain $1+1/\nu$ (and $2 - \eta$ where necessary) from Eq.~\eqref{QUOTIENTS}
with $O = \chi_{SG}$ and $O = \partial \xi_L / \partial \beta $
respectively.
\item
\label{stage3}
Using the value of $\omega$ from stage~\ref{stage1} and $1/\nu$ from
stage~\ref{stage2}, determine $\beta\crit$ from Eq.~\eqref{t*}.
\end{enumerate}
The error bars for $1+1/\nu$ and $2 - \eta$ from stage~\ref{stage2} will have a
systematic component, coming from the uncertainty in the value of $\omega$
from stage~\ref{stage1}, as well as a component from statistical errors in the
data being fitted. Similarly the error bar in $\beta\crit$ from
stage~\ref{stage3} will have a systematic component due to uncertainty in the
value of $\omega + 1/\nu$.

Each of these three stages only requires a straight-line fit. However, in
practice things are a little more tricky. We would like to use data for as
many sizes as possible, but in practice the smaller sizes are affected by
sub-leading corrections to scaling so we can only use data for the larger
sizes. It is therefore necessary to include only a range of sizes
for which the quality of the fit is satisfactory.

In some cases we try to incorporate a sub-leading correction to scaling to
increase the range of sizes that can be used.  These are of different types,
\textit{one} of which is higher powers of the leading correction, and this is
the only one we will include here in order to avoid introducing too many
additional parameters. In other words, when we include sub-leading corrections
we will do a \textit{parabolic}, rather than linear, fit to the data as a
function of $1/L^\omega$.

In order to increase the number of data points relative to the number of fit
parameters, we will often do a \textit{combined} fit to several data sets.
For example, when estimating $\omega$ we will determine the $\beta^\star_L$ from
one dimensionless quantity $f$, and then determine \textit{two}
(or more) other dimensionless quantities at these temperatures. 
These data sets will be simultaneously 
fitted to Eq.~\eqref{quotient_omega} with the \textit{same} value for
$\omega$ (since this is universal) but different amplitudes $B$ (since these
are non-universal). Hence, by combining two data sets,
we double the amount of data without doubling the
number of fit parameters. It should be mentioned that, for a given size, the
data for the different data sets is correlated, and best estimates of fitting
parameters are obtained by including these
correlations.\cite{ballesteros:96a,ballesteros:98,weigel:09} In other words,
if a data point is $(x_i, y_i)$, and the fitting function is $u(x)$, which
depends on certain fitting parameters, we determine those parameters by
minimizing
\begin{equation}
\chi^2 = \sum_{i, j} [y_i - u(x_i)]\, \left(C^{-1}\right)_{ij} \, 
[y_j - u(x_j)]\, , 
\end{equation}
where
\begin{equation}
C_{ij} = \langle y_i \, y_j \rangle -  \langle y_i \rangle \, \langle y_j \rangle \, ,
\end{equation}
is the covariance matrix.  If there  are substantial correlations in many
elements, the covariance matrix can become singular, but we have checked that
this is not the case for the quantities we study.

We end this section by discussing the FSS of a scale-invariant (but
dimensionfull) quantity, which turns out to be useful in our study of the LR
model. Take Eq.~\eqref{FSSleading} and imagine that we know exactly the
exponent $y_O$. Then, $O(L,t)/L^{y_O}$ is scale-invariant, since it remains
finite at $t=0$ even in the limit of large $L$. This is precisely the
situation in the LR model, if we take for $O$ the SG susceptibility, because,
as explained in the introduction, the anomalous dimension is a known function
of $\sigma$ for those models.  Nonetheless, Eq.~\eqref{FSS_nodimen} needs to
be modified when applied to $\chi_\text{SG}/L^{2\sigma-1}$, because the
magnetic scaling field $u(h, t)$ is not exactly $h$, as assumed in
Eq.~\eqref{fsing} (see
e.g. Refs.~\onlinecite{amit:05,hasenbusch:08b}). Rather, there is a non linear
dependency on the thermodynamic control parameters $t$ and $h$:
$u_h(h,t)=h\tilde u_h(t) + {\cal O}(h^3)$, where $\tilde u_h(t)= 1 +
c_1 t + c_2 t^2+ \cdots$.  Hence, the analogue of Eq.~\eqref{FSS_nodimen} reads
\begin{equation}\label{eq:fss-corrected}
\frac{\chi_\text{SG}(L,t)}{L^{2\sigma-1}}=\tilde u_h^2(t)\left[ \widetilde O_0(L^{1/\nu} t)
  + L^{-\omega}\widetilde O_1(L^{1/\nu} t) \right]\,.
\end{equation}
We note that the multiplicative renormalization $\tilde u_h^2(t)$ cancels out
when looking for crossing points, namely
\begin{equation}
\frac{\chi_\text{SG}(L,t^*_L)}{L^{2\sigma-1}}=
\frac{\chi_\text{SG}(sL,t^*_L)}{(sL)^{2\sigma-1}}\,,
\end{equation}
so $t^*_L$ scales as in Eq.~\eqref{t*}. Unfortunately, the
multiplicative renormalization can no longer be ignored when we compute $1/\nu$
from $\partial_\beta\chi_\text{SG}/L^{2\sigma-1}$.
Indeed, differentiating Eq.~\eqref{eq:fss-corrected}
with respect to $\beta$ and neglecting terms of order $1/L^{\omega+1/\nu}$, we find
\begin{eqnarray}
{\partial_\beta \chi_\text{SG}(L, t) \over L^{2\sigma - 1}} &=& L^{1/\nu} \,
\left[ \tilde u_h^2(t)\, \widetilde O'_0(L^{1/\nu} t) 
  \right. \\\nonumber &+& \left. L^{-\omega}\, 
  \tilde u_h^2(t)\,\widetilde O'_1(L^{1/\nu}t) \right. \\\nonumber &+& \left.
L^{-1/\nu}\, 2\tilde u_h(t)\tilde u_h'(t)\, \widetilde{O}_0(L^{1/\nu} t)\,
\right] \, ,
\label{chi_correction}
\end{eqnarray}
rather than Eq.~\eqref{FSSleading}.
Both $\tilde u_h$ and $\tilde u_h'$ behave as
$L$-independent constants (up to corrections of order $1/L^{\omega+1/\nu}$)
when evaluated at the crossing point $t^*_L$ given in Eq.~\eqref{t*}.
Hence, the quotient of the $\beta$ derivative of $\log \chi_\text{SG}$ is
given by
\begin{equation}
Q(\partial_\beta \log \chi_\text{SG}) = s^{1/\nu} + B_1 L^{-\omega} + B_2
L^{-1/\nu} \, ,
\label{quotient_chi}
\end{equation}
instead of Eq.~\eqref{QUOTIENTS}, showing that 
there are corrections of order $L^{-1/\nu}$
as well as $L^{-\omega}$. For
some values of $\sigma$, and also the 3-$d$ SR model,\cite{hasenbusch:08b}
one finds $1/\nu < \omega$ so
the $L^{-1/\nu}$ correction dominates. 

\section{Simulation Details}
\label{sec:simulation}

For each size and temperature we simulate four copies of the spins with the same
interactions. By simulating four copies we can calculate, without bias, quantities
which involve a product of up to four thermal averages, such as
the spin glass susceptibility, Eq.~\eqref{chiSG}, 
the $U_4$ moment ratio, \eqref{U4}, and derivatives of these
quantities with respect to $\beta$ calculated from Eq.~\eqref{beta_deriv}.

The simulations use parallel tempering\cite{hukushima:96} (PT) to speed up
equilibration. For the same set of interactions we study $N_\beta$ values of
$\beta$ between $\beta_\text{max}$ and $\beta_\text{min}$. To obtain good
statistics we simulate a large number,
$N_\text{samp}$, of samples, where $N_\text{samp}$ is a
multiple of 128 because 128 samples are simulated in parallel by multispin
coding. For the long-range models there are $N_\text{samp}/128$ different graphs, but
each sample for the same graph has different interactions.
We run for $N_\text{sweep}$ single-spin flip (Metropolis) sweeps performing a
parallel tempering sweep every 10 Metropolis sweeps. 
The parameters used for the different models are shown in
Tables~\ref{tab:params4d}--\ref{tab:params896}.

\begin{table}[!tb]
\caption{
Parameters of the simulations of the 4-$d$ model:
$N_\beta$ is the  number of temperatures with $\beta_\text{max}$ the largest
and  $\beta_\text{min}$ the smallest. The number of Metropolis sweeps is given
by $N_\text{sweep}$, and the number of samples is $N_\text{samp}$.
}
\label{tab:params4d}
\begin{tabular*}{\columnwidth}{@{\extracolsep{\fill}}rrcccc}
\hline
\hline
\multicolumn{1}{c}{$L$} & $N_\text{sweep}$ & $N_\beta$ & $\beta_\text{max}$ & $\beta_\text{min}$ & $N_\text{samp}$\\
\hline
4   & $2.56\times10^5$ & 23        & 0.5025             & 0.4                & $2^{20}$ \\
5   & $2.56\times10^5$ & 23        & 0.5025             & 0.4                & $2^{20}$ \\
6   & $2.56\times10^5$ & 23        & 0.5025             & 0.4                & $2^{20}$ \\
8   & $2.56\times10^5$ & 23        & 0.5025             & 0.4                & $2^{20}$ \\
10  & $2.56\times10^5$ & 23        & 0.5025             & 0.4                & $2^{20}$ \\
12  & $2.56\times10^5$ & 23        & 0.5025             & 0.4                & $2^{20}$ \\
16  & $5.12\times10^5$ & 23        & 0.5025             & 0.4                & $2^{20}$ \\
\hline
\hline
\end{tabular*}
\end{table}

\begin{table}[!tb]
\caption{
Parameters of the simulations of the 1-$d$ model with $\sigma = 0.790$. See
Table~\ref{tab:params4d} for an explanation of the symbols.
}
\label{tab:params790}
\begin{tabular*}{\columnwidth}{@{\extracolsep{\fill}}rrcccc}
\hline
\hline
\multicolumn{1}{c}{$L$}  & $N_\text{sweep}$ & $N_\beta$ & $\beta_\text{max}$ & $\beta_\text{min}$ & $N_\text{samp}$\\
\hline
512   & $10^6$           & 16        & 0.671             & 0.538               & 64000  \\
1024  & $10^6$           & 16        & 0.671             & 0.538               & 64000  \\
2048  & $10^6$           & 16        & 0.671             & 0.538               & 64000  \\
4096  & $1.28\times10^6$ & 16        & 0.671             & 0.538               & 64000  \\
8192  & $1.28\times10^6$ & 16        & 0.671             & 0.538               & 64000  \\
16384 & $2 \times 10^6$  & 16        & 0.671             & 0.538               & 64000  \\
32768 & $2 \times 10^6$  & 16        & 0.671             & 0.538               & 64000  \\
\hline
\hline
\end{tabular*}
\end{table}

\begin{table}
\caption{
Parameters of the simulations of the 1-$d$ model with $\sigma = 0.896$. See
Table~\ref{tab:params4d} for an explanation of the symbols. 
}
\label{tab:params896}
\begin{tabular*}{\columnwidth}{@{\extracolsep{\fill}}rrcccc}
\hline
\hline
\multicolumn{1}{c}{$L$}  & $N_\text{sweep}$   & $N_\beta$ & $\beta_\text{max}$ & $\beta_\text{min}$ & $N_\text{samp}$\\
\hline
512   & $1.28\times 10^6$ & 16        & 1.5                & 0.6                & 12800  \\
1024  & $2.56\times 10^6$ & 13        & 1.2                & 0.6                & 12800  \\
2048  & $1.024\times 10^7$& 14        & 1.2                & 0.65               & 12800  \\
4096  & $8.192\times 10^7$& 16        & 1.2                & 0.65               & 12800  \\
8192  & $8.192\times 10^7$& 16        & 1.1                & 0.71               & 12800  \\
\hline
\hline
\end{tabular*}
\end{table}

To check that the simulations were run for long enough to ensure equilibration we
adopted the following procedure. We divide the measurements into bins whose
size varies logarithmically, the first averages over the last half
of the sweeps, i.e.\ between sweeps $N_\text{sweep}$ and $N_\text{sweep}/2$, the
second averages between sweeps $N_\text{sweep}/2$ and $N_\text{sweep}/4$, the
third between sweeps $N_\text{sweep}/4$ and $N_\text{sweep}/8$, etc. We
require that the difference between the results in the first two bins is zero
within the error bars, where we get the error bar for the difference by
forming the difference 
between the results for the two bins separately for each sample before
averaging over samples. In most cases, to be on the safe side, we
actually require
that the differences between the first \textit{three} bins are all zero within errors.

\begin{figure}
\includegraphics[width=\columnwidth]{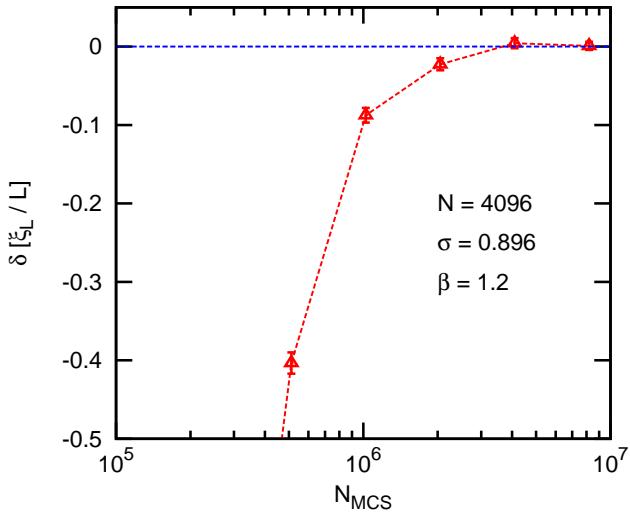}
\caption{(Color online)
The difference in the value of the $\xi_L/L$ between measurements obtained in the
range of sweeps $N_\text{MCS}/2$ to $N_\text{MCS}$
and measurements in the range $N_\text{MCS}/4$ to $N_\text{MCS}/2$, for
values of $N_\text{MCS}$ increasing by factors of 2 up to $N_\text{sweep}
= 8.192 \times
10^7$. The data is for the long-range model with $\sigma = 0.896$ at $\beta =
1.2$, the lowest temperature studied.
}
\label{fig:equil}
\end{figure}

This procedure
is illustrated in Fig.~\ref{fig:equil} which shows data for the
long-range model with $V = 4096, \sigma = 0.896$ at $\beta = 1.2$, the largest
$\beta$ value that we studied.
The vertical axis is the difference in $\xi_L/L$ between the bin containing
measurements in sweeps $N_\text{MCS}/2$ to $N_\text{MCS}$ and the bin for
sweeps in the interval $N_\text{MCS}/4$ to $N_\text{MCS}/2$, for different values
of $N_\text{MCS}$ up to $N_\text{sweep} = 8.192 \times 10^7$, the value in
Table \ref{tab:params896}. Since the two points for the largest number of
sweeps are zero within errors, it follows that the first \textit{three} bins
all agree.

\section{Results}
\label{sec:results}

\subsection{Four-dimensional short range model}
\label{sec:res-4d}
\begin{figure}
\includegraphics[width=\columnwidth]{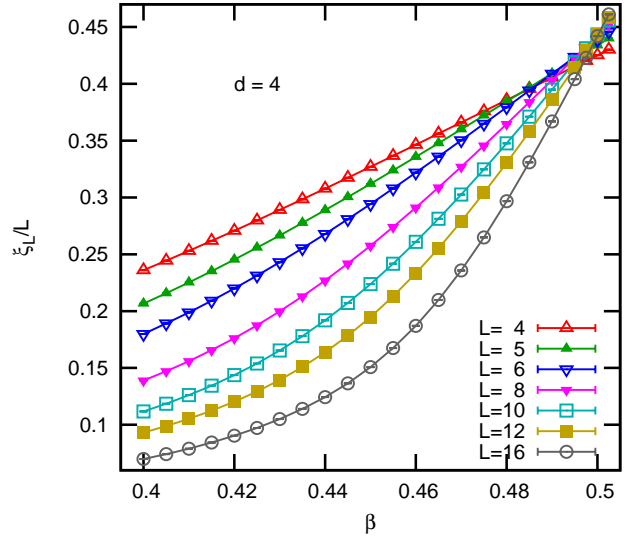}
\caption{(Color online)
A global view of the
data for the correlation length divided by $L$ for the 4-$d$ model.
}
\label{fig:xi_4d}
\end{figure}

\begin{figure}
\includegraphics[width=\columnwidth]{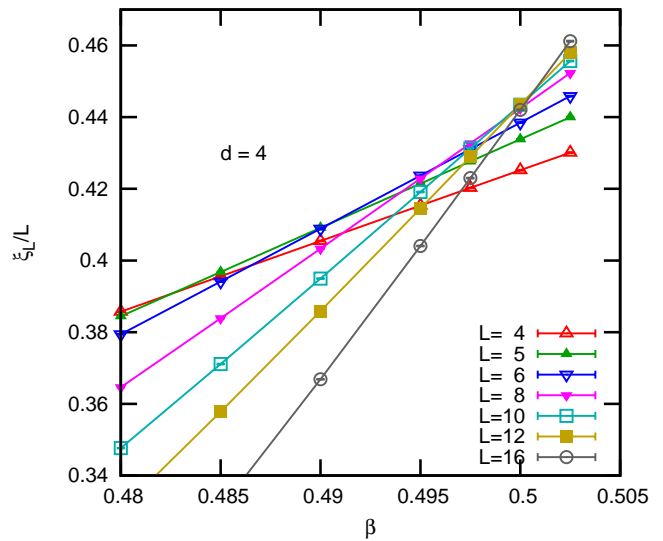}
\caption{(Color online)
An enlarged view of the data in Fig.~\ref{fig:xi_4d} showing the region of the
intersections.
}
\label{fig:xi_4d_blowup}
\end{figure}

\begin{figure}
\includegraphics[width=\columnwidth]{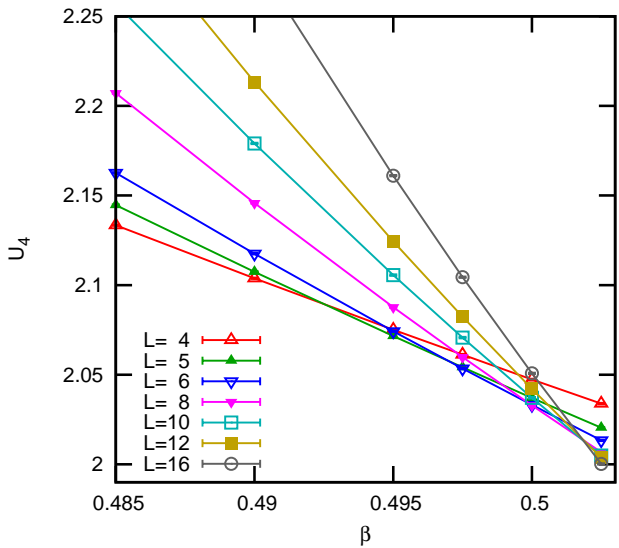}
\caption{(Color online)
An enlarged view of the data for $U_4$ for the 4-$d$ model
showing the region of the intersections.
}
\label{fig:U4_4d_blowup}
\end{figure}

Figures \ref{fig:xi_4d} and \ref{fig:xi_4d_blowup} show results for $\xi_L/L$
defined in Eq.~\eqref{xiL}
and Fig.~\ref{fig:U4_4d_blowup} shows results for the dimensionless ratio of
moments $U_4$ defined in Eq.~\eqref{U4}.
The resulting inverse temperatures $\beta^\star_L$  where data for sizes $L$ and $2L$
intersect, i.e.\ where their quotient $Q$ is unity,
is shown in Table~\ref{tab:4d_T*}. Results are given for both
$\xi_L/L$ and $U_4$.

\begin{table}
\caption{
Inverse temperatures $\beta^\star_L$, where data for sizes $L$ and $2L$ intersect,
i.e.\ where the quotient $Q$ is equal to unity,
for $\xi_L/L$ and the ratio of moments $U_4$,
for the 4-$d$ short-range model.
}
\label{tab:4d_T*}
\begin{tabular*}{\columnwidth}{@{\extracolsep{\fill}}rcc}
\hline
\hline
$L$ & $\beta^\star_L$ where $Q(\xi_L/L) = 1$ & $\beta^\star_L$ where $Q(U_4) = 1$ \\
\hline
4  &  $0.49113 \pm 0.00009$  & $0.49725 \pm 0.00011$ \\
5  &  $0.49598 \pm 0.00007$  & $0.50001 \pm 0.00009$ \\
6  &  $0.49825 \pm 0.00006$  & $0.50118 \pm 0.00008$ \\
8  &  $0.50012 \pm 0.00005$  & $0.50180 \pm 0.00006$ \\
\hline
\hline
\end{tabular*}
\end{table}

\begin{figure}
\includegraphics[width=\columnwidth]{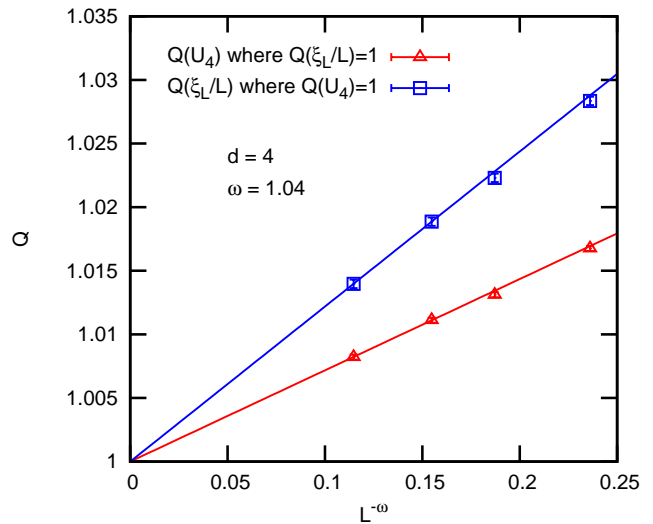}
\caption{(Color online)
The quotient of the dimensionless quantity $\xi_L/L$ of the 4-$d$ model
at the $U_4$ crossing (squares) and the quotient of
$U_4$ at the $\xi_L/L$ crossing (triangles). The straight lines represent the best
fit to Eq.~\eqref{quotient_omega} using the largest two sizes,
with the correction to scaling exponent $\omega$ as an
adjustable parameter.
}
\label{fig:Q_omega_4d}
\end{figure}

To compute the correction to scaling exponent $\omega$ we determine the
quotient of $\xi_L/L$ at the $U_4$ crossing and vice versa. These quotients
are shown in Table \ref{tab:4d_quotients} and plotted in
Fig.~\ref{fig:Q_omega_4d}. Fitting the largest two
pairs of sizes for each quantity to
Eq.~\eqref{quotient_omega} for $s=2$ with the same exponent $\omega$ gives 
\begin{equation} 
\omega_\mathrm{SR}(4)=1.04(10), \qquad \chi^2/\text{dof}=0.99/1 \, .
\label{omega_4d}
\end{equation}
It should be mentioned that the lines in Fig.~\ref{fig:Q_omega_4d} are
not separate fits to each set of data but are combined fits including the whole
covariance matrix.

We have tried also fits including subleading corrections to scaling. For instance,
considering, in addition, the quotient of
$R_{12}$, defined in Eq.~\eqref{R12},
at the crossings of $\xi_L/L$ and $U_4$, and fitting the three largest
sizes to $1+B_1 L^{-\omega} + B_2 L^{-2\omega}$ gives a satisfactory fit with 
$\omega=1.29(26), \chi^2/\text{dof}=2.26/5$. However we prefer the result
$\omega=1.04(10)$ since it has been obtained using larger
lattices $(L \ge 6)$.

\begin{table}
\caption{
Quotients of $U_4$ at the crossings of $\xi_L/L$, and vice versa, for the
4-$d$ short-range model.
}
\label{tab:4d_quotients}
\begin{tabular*}{\columnwidth}{@{\extracolsep{\fill}}rcc}
\hline
\hline
$L$ & $Q(U_4)$ where $Q(\xi_L/L) = 1$; & $Q(\xi_L/L)$ where $Q(U_4) = 1$ \\
\hline
4  &  $1.01675 \pm 0.00020$  & $1.02835 \pm 0.00033$ \\
5  &  $1.01311 \pm 0.00020$  & $1.02230 \pm 0.00033$ \\
6  &  $1.01112 \pm 0.00020$  & $1.01886 \pm 0.00033$ \\
8  &  $1.00822 \pm 0.00020$  & $1.01397 \pm 0.00033$ \\
\hline
\hline
\end{tabular*}
\end{table}

\begin{figure}
\includegraphics[width=\columnwidth]{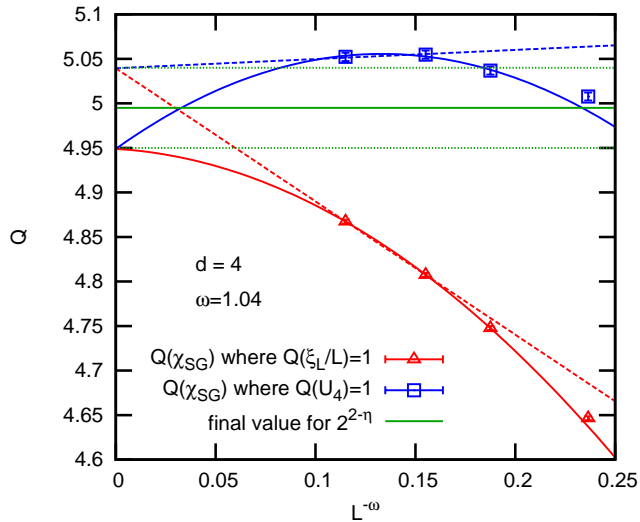}
\caption{(Color online)
The quotients of $\chi_{SG}$ of the 4-$d$ model
at the crossings of $\xi_L/L$ (triangles) and
$U_4$ (squares) as a function of $L^{-\omega}$, where $\omega$ has already been
determined, see Fig.~\ref{fig:Q_omega_4d}, and is given by
Eq.~\eqref{omega_4d}. The dashed lines are the linear fit,
with a common intercept on the $y$ axis,
to the two largest pairs of sizes,
and the solid lines are the quadratic fit to the three largest pairs of
sizes (again with a common intercept). The intercept is equal to $2^{2-\eta}$.
The horizontal lines indicate the final estimate and error bars for $Q$
given in
Eq.~\eqref{Q_eta}. This leads to the final estimate for $\eta$ in
Eq.~\eqref{eta_4d}.
}
\label{fig:Q_eta_4d}
\end{figure}

Next we compute $\eta$ from the quotients of $\chi_{SG}$, defined in
Eq.~\eqref{chiSG}, at the crossings of
$\xi_L/L$ and $U_4$, which are shown in Table \ref{tab:4d_Q_eta} and Figures
\ref{fig:xi_4d_blowup} and \ref{fig:U4_4d_blowup}.
Assuming $\omega=1.04(10)$, a linear fit to
Eq.~\eqref{QUOTIENTS} with $s=2$ and the same value of $y_O\, (=2-\eta)$ for
both quantities gives,
for the largest two pairs of sizes, 
$Q \equiv 2^{2-\eta} =4.949(45)[^{+8}_{-14}], \chi^2/\text{dof}=0.42/1,$
in which the numbers in rectangular brackets, $[\cdots]$,
correspond to the errors due to the uncertainty in the value of $\omega$.
This fit is shown in Fig.~\ref{fig:Q_eta_4d} by the dashed lines.

On the other hand, a quadratic fit to
$Q(\chi_{SG})=Q+B_1 L^{-\omega}+B_2 L^{-2\omega}$ using the largest three pairs
gives $ Q=5.039(10)[^{+20}_{-16}],\ \chi^2/\text{dof}=0.076/1$, which is also an
acceptable fit, shown by the solid lines in Fig,~\ref{fig:Q_eta_4d}.

If we assume the larger value for $\omega$ discussed above,
namely $\omega=1.29(26)$ we
find that only a quadratic fit is acceptable, and the value for $Q$ is
$Q=4.962(30)[6], \chi^2/\text{dof}=0.011/1$,
which is intermediate between the two previous values of $Q$. We can
summarize all the numbers with the value
\begin{equation}
Q \equiv 2^{2-\eta} = 4.994(45).
\label{Q_eta}
\end{equation}
The central value is shown as the solid horizontal line in
Fig.~\ref{fig:Q_eta_4d}, and the error bars are indicated by the dotted
horizontal lines.
Equation \eqref{Q_eta} gives
\begin{equation}
\eta_\mathrm{SR}(4) = -0.320(13) \, .
\label{eta_4d}
\end{equation}

\begin{table}
\caption{
Quotients of $\chi_{SG}$ at the crossings of $\xi_L/L$ and $U_4$ for the 4-$d$
short-range model.
}
\label{tab:4d_Q_eta}
\begin{tabular*}{\columnwidth}{@{\extracolsep{\fill}}rcc}
\hline
\hline
$L$ & $Q(\chi_{SG})$ where $Q(\xi_L/L) = 1$; & $Q(\chi_{SG})$ where $Q(U_4) = 1$ \\
\hline
4  &  $4.6464 \pm 0.0022$   &   $5.0077 \pm 0.0045$ \\
5  &  $4.7477 \pm 0.0022$   &   $5.0368 \pm 0.0046$ \\
6  &  $4.8074 \pm 0.0022$   &   $5.0547 \pm 0.0047$ \\
8  &  $4.8673 \pm 0.0022$   &   $5.0522 \pm 0.0047$ \\
\hline
\hline
\end{tabular*}
\end{table}

\begin{table}
\caption{
Quotients of the $\beta$ derivative of $\xi_L$ at the crossings of $\xi_L/L$
for the 4-$d$ short-range model.
}
\label{tab:4d_Q_nu}
\begin{tabular*}{\columnwidth}{@{\extracolsep{\fill}}rc}
\hline
\hline
$L$ & $Q(\partial_\beta \xi_L)$ where $Q(\xi_L/L) = 1$ \\
\hline
4  & $3.9581 \pm 0.0024$\\
5  & $3.9340 \pm 0.0026$\\
6  & $3.9133 \pm 0.0025$\\
8  & $3.8936 \pm 0.0031$\\
\hline
\hline
\end{tabular*}
\end{table}

\begin{figure}
\includegraphics[width=\columnwidth]{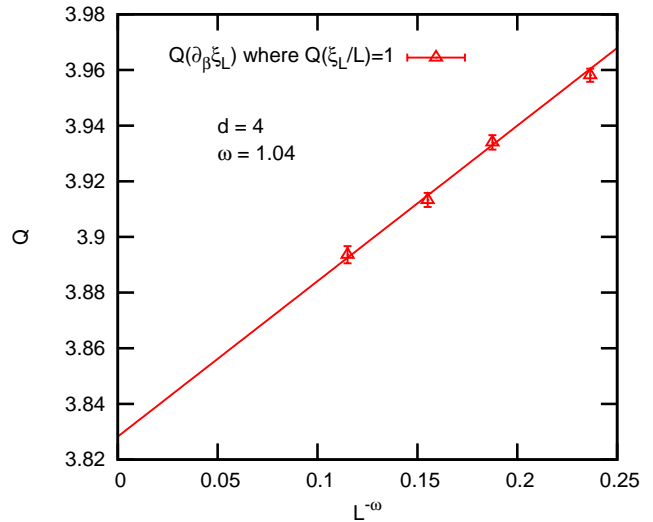}
\caption{(Color online)
The quotient of $\partial_\beta \xi_L$ of the 4-$d$ model
at the crossing of $\xi_L/L$.
as a function of $L^{-\omega}$, where $\omega$ has already been
determined, see Fig,~\ref{fig:Q_omega_4d}, and is given by
Eq.~\eqref{omega_4d}. The solid line is a linear fit to  Eq.~\eqref{QUOTIENTS}
using the three largest pairs of sizes.
The intercept is equal to $2^{1+1/\nu}$. The final value of $\nu$ is given in
Eq.~\eqref{nu_4d}.
}
\label{fig:Q_nu_4d}
\end{figure}

To compute $\nu$ we have used the quotients for the
$\beta$-derivative of $\xi$ at the crossings of $\xi_L/L$.
The values for each pair are given in Table \ref{tab:4d_Q_nu}.
Taking $\omega=1.04(10)$
we obtain, fitting the three largest pairs, to Eq.~\eqref{QUOTIENTS} for
$s=2$,
\begin{equation}
Q \equiv 2^{1 + 1/\nu} = 3.828(9)[8],\qquad \chi^2/\text{dof}=0.68/1 \, ,
\end{equation}
which gives $\nu=1.068(4)[3].$ Combining the errors we get our final estimate
for $\nu$ as 
\begin{equation}
\nu_\mathrm{SR}(4) = 1.068(7) \, .
\label{nu_4d}
\end{equation}
The data and the fit are shown in Fig.~\ref{fig:Q_nu_4d}

\begin{figure}
\includegraphics[width=\columnwidth]{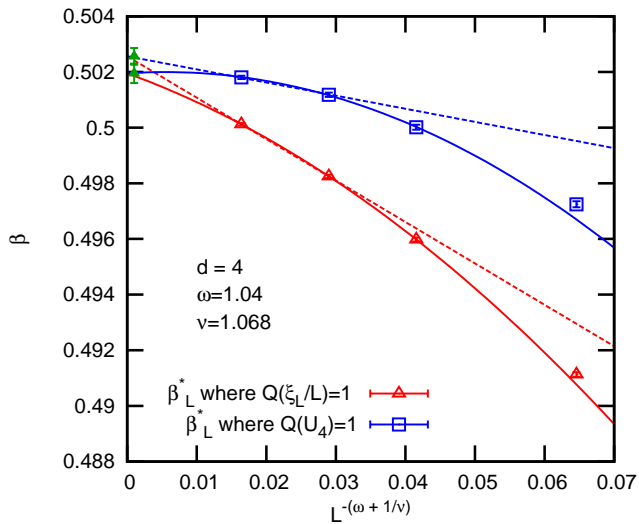}
\caption{(Color online)
Values of $\beta^\star_L$,
the crossing points for $\xi_L/L$ and $U_4$,
for the 4-$d$ model,
and fits as a function of
$1/L^{\omega + 1/\nu}$, for the 4-$d$ model. We used the values of $\omega$
and
$\nu$ previously determined, see Eqs.~\eqref{omega_4d} and \eqref{nu_4d}.
The dashed lines are the linear fit, according to Eq.~\eqref{t*},
with a common intercept on the $y$ axis,
to the two largest pairs of sizes,
and the solid lines are the quadratic fit to the three largest pairs of
sizes (again with a common intercept). The intercept is the critical coupling
$\beta\crit$.
The green data points are the estimates for $\beta\crit$ for the two fits,
Eqs.~\eqref{beta1} and \eqref{beta2}.
}
\label{fig:betac_4d}
\end{figure}

Finally we estimate
$\beta\crit$ by fitting the crossing points for $\xi_L/L$ and $U_4$ to
Eq.~\eqref{t*}, using the previously determined values $\omega=1.04(10)$ and
$\nu=1.068(7)$.
The data has already been given in Table \ref{tab:4d_T*} and is
plotted in Fig.~\ref{fig:betac_4d}.
We obtain a good fit 
considering only the (6,12) and (8,16)
pairs:
\begin{equation}
\beta\crit = 0.50256(14)[15], \qquad \chi^2/\text{dof} = 0.24/1 \, .
\label{beta1}
\end{equation}
This fit is shown by the dashed lines in Fig.~\ref{fig:betac_4d}.

We have tried to (roughly) take into account higher order corrections to
scaling adding
a quadratic term in $L^{-\omega-1/\nu}$. We obtain a good fit with the pairs
(5,10), (6,12) and (8,16):
\begin{equation}
\beta\crit = 0.50195(34)[1], \qquad \chi^2/\text{dof}=0.30/1 \, ,
\label{beta2}
\end{equation}
and this is shown by the solid lines in Fig.~\ref{fig:betac_4d}.
We can therefore safely take the value,
\begin{equation}
\beta\crit = 0.5023(6) \ \Rightarrow \ T\crit = 1.9908(24) \quad (d=4)\, ,
\end{equation}
as our final result.

We end this section by comparing our results with previous computations
by other authors. Marinari and Zuliani\cite{marinari:99} studied the
4-$d$ spin glass with binary couplings, finding $T\crit=2.03(3)$,
$\nu=1.00(10)$ and $\eta=-0.30(5)$, in good agreement with our more
accurate estimates. J\"org and Katzgraber\cite{jorg:08c} studied a different
version of the
4-$d$ spin glass which is expected to belong to the same
universality class. They found $\nu=1.02(2)$ and $\eta=-0.275(25)$, which are two
standard deviations from our estimate. J\"org and Katzgraber also
considered the leading corrections to scaling, but found an extremely large
exponent, $\omega\approx 2.5$. They were aware that such a
large $\omega$ is unlikely to be correct, 
and they attributed their result to the small lattice sizes
that they could equilibrate.

\subsection{One-dimensional long range model with $\boldsymbol{\sigma =0.790}$}
\label{sec:res-790}

\begin{figure}
\includegraphics[width=\columnwidth]{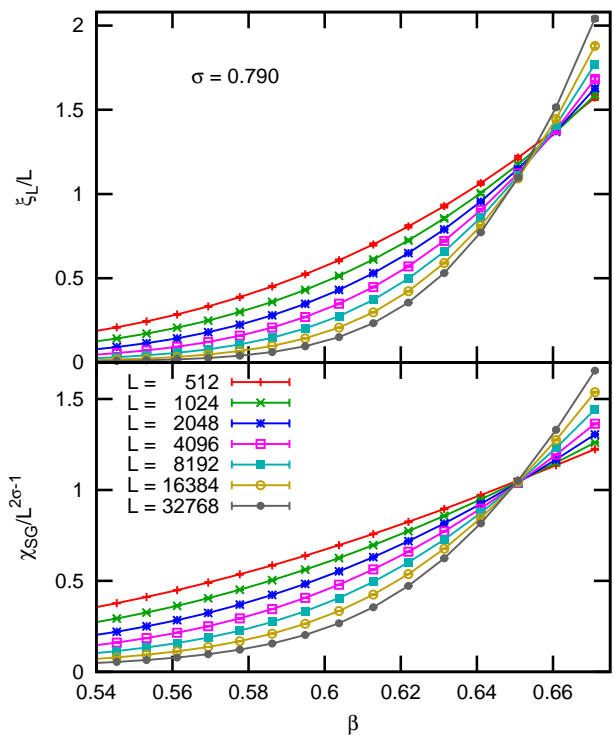}
\caption{(Color online) Correlation length in units of the system size ({\bf
top}) and scale-invariant combination of the SG susceptibility and the
lattice dimension $\chi_\text{SG}/L^{2\sigma-1}$ ({\bf bottom}), as a function of the
inverse temperature $\beta$, for the LR-model with
$\sigma=0.790$. For both quantities, the curves
for the different $L$ should cross at temperatures that approach the
critical point when $L$ grows, see Eq.~\eqref{t*}.  }
\label{fig:scaleinvariant_790}
\end{figure}

From Eq.~\eqref{sigma_eta} and the value $\eta_\mathrm{SR}(4) = -0.320(13)$
for the 4-$d$ model given in
Eq.~\eqref{eta_4d}, we see that
$\sigma = 0.790$ is a proxy for the 4-$d$ short-range model, at least according
to the comparison of the exponents $\eta$ (or equivalently of the magnetic
exponents $y_H$, see Eq.~\eqref{SR_LR}).
In this section we will see if Eq.~\eqref{SR_LR} is also
satisfied for the thermal exponents $y_T$ (for which Eq.~\eqref{SR_LR} can be
expressed in terms of $\nu$ as shown in Eq.~\eqref{nu_compare}), and the correction to
scaling exponents $\omega\, (= - y_u)$.
Since 
$\eta_\mathrm{LR}$ is known exactly, $2-\eta_\mathrm{LR}(\sigma) = 2\sigma - 1$, see
Eq.~\eqref{eta_LR}, we can include
$\chi_{SG}/L^{2\sigma - 1}$ as another scale invariant quantity to be studied.

\begin{table}
\caption{
Inverse temperatures $\beta^\star_L$, where data for sizes $L$ and $2L$
intersect,
i.e.\ where the quotient $Q$ is equal to unity,
for $\chi_{SG}/L^{2\sigma-1}$ and $\xi_L/L$
for the LR model with $\sigma = 0.790$.
}
\label{tab:790_T*}
\begin{tabular*}{\columnwidth}{@{\extracolsep{\fill}}rcc}
\hline
\hline
\multicolumn{1}{c}{$L$} &$\beta^\star_L$ where $Q(\chi_{SG}/L^{2\sigma - 1})\!=\! 1$ &
$\beta^\star_L$ where $Q(\xi_L/L)\!=\! 1$ \\
\hline
512   & $0.6538 \pm 0.0020$ & $0.6665 \pm 0.0066$ \\
1024  & $0.6532 \pm 0.0018$ & $0.6598 \pm 0.0050$ \\
2048  & $0.6516 \pm 0.0014$ & $0.6586 \pm 0.0038$ \\
4096  & $0.6498 \pm 0.0012$ & $0.6545 \pm 0.0031$ \\
8192  & $0.6500 \pm 0.0009$ & $0.6541 \pm 0.0023$ \\
16384 & $0.6492 \pm 0.0008$ & $0.6501 \pm 0.0019$ \\
\hline
\hline
\end{tabular*}
\end{table}

\begin{figure}
\includegraphics[width=\columnwidth]{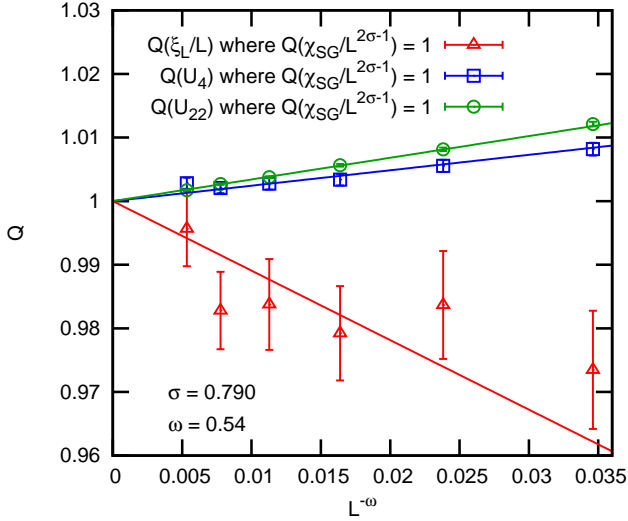}
\caption{(Color online)
The quotients of dimensionless quantities $\xi_L/L, U_4$ and $U_{22}$ for
$\sigma = 0.790$ at the
crossing of $\chi_{SG}/L^{2\sigma-1}$.
The straight lines represent the
best
fit to Eq.~\eqref{quotient_omega} using all the data,
with the correction to scaling exponent $\omega$ as an
adjustable parameter.
}
\label{fig:Q_omega_790}
\end{figure}

\begin{figure}
\includegraphics[width=\columnwidth]{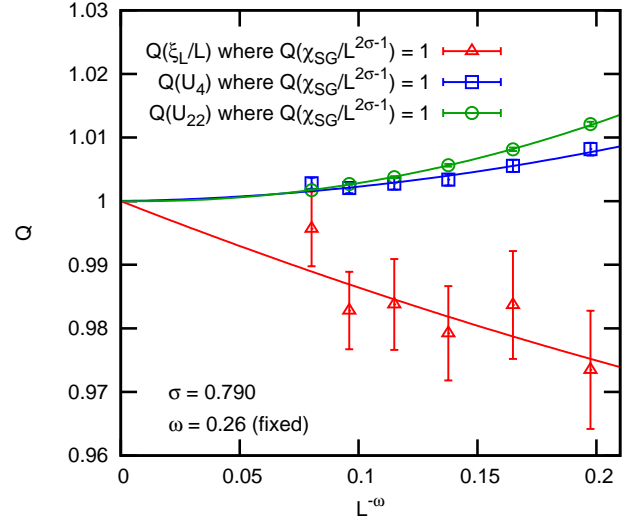}
\caption{(Color online)
The quotients of dimensionless quantities
$\xi_L/L, U_4$ and $U_{22}$ at the
crossings of $\chi_{SG}/L^{2\sigma-1}$ for $\sigma = 0.790$.
The lines represent the
best quadratic fit as function of $1/L^\omega$,
using all the data,
where $\omega$ is fixed at $0.26 \ (=1.04/4)$, the value expected from
the correspondence with the 4-$d$ model, for which the value of $\omega$ is
given in Eq.~\eqref{omega_4d}.
}
\label{fig:Q_omega_790_quad}
\end{figure}

We focus on $\xi_L$ and $\chi_{SG}/L^{2\sigma -1}$, data for which are shown in
Fig.~\ref{fig:scaleinvariant_790}, and the corresponding
crossing points
are given in Table \ref{tab:790_T*}.  Our first task is to try to
determine the correction to scaling exponent $\omega$. We fit the
quotients of $\xi_L/L$, $U_4$, and $U_{22}$ defined in Eq.~\eqref{U22}, at the
crossing of $\chi_{SG}/L^{2\sigma-1}$, including all
the $(L,2L)$ pairs. A straight line fit, shown in Fig.~\ref{fig:Q_omega_790},
is acceptable:
\begin{equation}
\omega=0.539(9),\qquad \chi^2/\text{dof}=16.7/14 ,
\end{equation}
and has a probability of 15\%. A quadratic fit to
$1+B_1 L^{-\omega} + B_2 L^{-2 \omega}$ gives a better fit:
$\omega=0.29(-4+9), \chi^2/\text{dof}=7/11$. This is consistent with the 
value $0.26(3)$ expected from the correspondence in Eq.~\eqref{omega_compare}
and the value 
of $\omega$ for the 4-$d$ model given in Eq.~\eqref{omega_4d}. We have also
tried fits in which $\omega$ is \textit{fixed} to the value $0.26$. A straight line fit using
all the data is very poor, $\chi^2/\text{dof}= 1069/15$, whereas a quadratic
fit works well, $\chi^2/\text{dof}=7.5/12$, and is shown in
Fig.~\ref{fig:Q_omega_790_quad}.

Altogether, we see that our data for the quotients of scale invariant quantities
do not constrain $\omega$ precisely. Any
value in the range $0.25$--$0.55$ can be considered acceptable. Fortunately,
this includes the value expected from the the correspondence with the 4-$d$
model, $\omega = 0.26(3)$.

\begin{figure}
\includegraphics[width=\columnwidth]{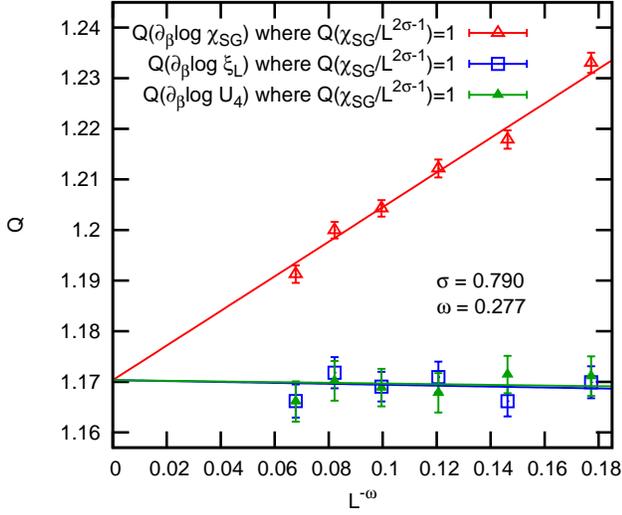}
\caption{(Color online)
The quotients of $\partial_\beta \log\xi_L, \partial_\beta \log U_4$ and
$\partial_\beta \log \chi_\text{SG}$ at the
crossings of $\chi_{SG}/L^{2\sigma-1}$ for $\sigma = 0.790$.
The lines represent the
best straight-line fit as function of $1/L^\omega$,
using all the data,
in which $\omega$, as well as the intercept $Q = 2^{1/\nu}$, is a fit
parameter. 
}
\label{fig:Q_nu_790}
\end{figure}

To estimate $\nu$ we consider the $(L, 2L)$ quotients of the
logarithmic derivative of
$\chi_\text{SG}, \xi_L$, and $U_4$ with respect to $\beta$,
at the crossings of $\chi_{SG}/L^{2\sigma -1}$. 
All these quotients should tend to $2^{1/\nu}$ for $L \to \infty$. A
straight-line fit according to Eq.~\eqref{QUOTIENTS}, allowing $\omega$ as well as the
intercept $Q$ to vary, is shown in Fig.~\ref{fig:Q_nu_790}. The
result is 
\begin{align}
&Q \equiv 2^{1/\nu} = 1.1703(23),\quad \chi^2/\text{dof}=14.24/13 ,  \\
&\omega_\mathrm{LR}(0.790) = 0.277(8) \, , 
\label{omega_790}
\end{align}
which gives
\begin{equation}
\nu_\mathrm{LR}(0.790) = 4.41(19)\, .
\label{nu_790}
\end{equation}
This is consistent with the result $4.272(20)$ expected from the
correspondence with the 4-$d$ model, see Eq.~\eqref{nu_compare}, and the the
4-$d$ value of $\nu$ given in Eq.~\eqref{nu_4d}, $\nu_\mathrm{SR}(4) =
1.068(7)$.  It is surprising that the fits in Fig.~\ref{fig:Q_nu_790} gives
such a good precision for $\omega$, better than using quotients of scale
invariant quantities which we showed in Figs.~\ref{fig:Q_omega_790} and
\ref{fig:Q_omega_790_quad}.  The result $\omega = 0.277(8)$ is consistent with
that expected from the 4-$d$ correspondence, $\omega = 0.26(3)$. We have also
tried a quadratic fit, which gives $Q=1.1742(58)[22],
\chi^2/\text{dof}=9.54/11$, and a linear fit discarding the $L=512$ data which
gives $Q=1.1683(15)[62], \chi^2/\text{dof}=7.56/8$ (both of these fits used
the value for $\omega$ obtained from the correspondence with the 4-$d$ model,
$\omega = \omega_\mathrm{SR}(4)/4 = 0.26(3)$). These results are all
consistent with Eq.~\eqref{nu_790} which we therefore take as 
our final estimate for $\nu_\mathrm{LR}(0.790)$. 

However, the alert reader
will recall from Sec.~\ref{sec:fss}
that the $\beta$-derivative of $\chi_\text{SG}/L^{2\sigma-1}$ suffers from
\textit{two} types of corrections to scaling, one of order $L^{-\omega}$ and the other
of order $L^{-1/\nu}$, see Eqs.~\eqref{chi_correction} and \eqref{quotient_chi}.
The relationship between LR and SR exponents
in Eqs.~\eqref{nu_compare} and \eqref{omega_compare}, combined
with our numerical results for the $d=4$ SR-model in Sect.~\ref{sec:res-4d}, suggests that
the two corrections to scaling are very similar for
$\sigma = 0.790$ because $\omega_\text{SR}(4)\simeq 1/\nu_\text{SR}(4)$.
This implies that the two corrections
can be lumped together into a single term to a good
approximation.
Indeed, we have succeeded in analyzing our numerical data by considering
only the scaling corrections of order $L^{-\omega}$.
Therefore, although we take Eq.~\eqref{omega_790} as our
final estimate for $\omega_\text{LR}(0.790)$, we warn that its error is probably
underestimated, due to the oversimplification in the functional form for the
scaling corrections.

By contrast, we
shall see in Sect.~\ref{sec:res-896} that for $\sigma = 0.896$ the corrections of order
$L^{-1/\nu}$ turn out to be dominant, and will need to be taken into
account explicitly.

\begin{figure}
\includegraphics[width=\columnwidth]{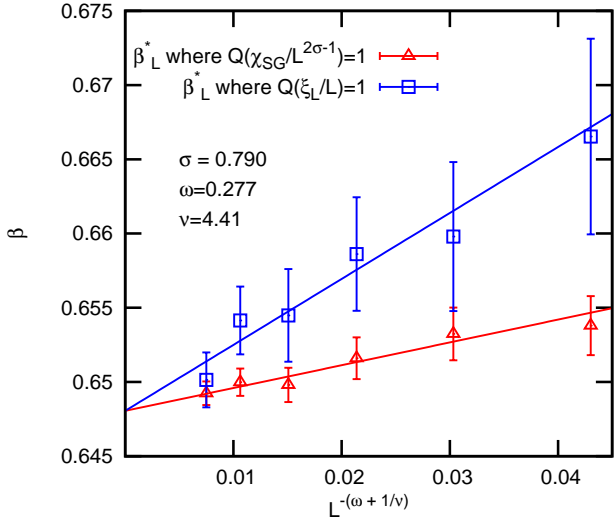}
\caption{(Color online)
Values of $\beta^\star_L$, the crossing points for $\xi_L/L$ and
$\chi_{SG}/L^{2\sigma-1}$, for $\sigma = 0.790$,
as a function of $1/L^{\omega + 1/\nu}$ where the
values of $\omega$ and $\nu$ are fixed at the values given in
Eqs.~\eqref{omega_790}
and \eqref{nu_790}. The intercept is
the critical coupling $\beta\crit$.
}
\label{fig:betac_790}
\end{figure}

Finally, in this section, we determine $\beta\crit$ by fitting
the crossing points of $\xi_L/L$ and
$\chi_{SG}/L^{2\sigma-1}$ shown in Table~\ref{tab:790_T*}
to Eq.~\eqref{t*}, assuming the values in Eq.~\eqref{omega_790} and
\eqref{nu_790}, 
$\omega=0.277(8), \nu=4.41(19)$. The plot is shown in
Fig.~\ref{fig:betac_790}, and the result is $\beta\crit = 0.64805(39)[2]$.
Combining the errors gives
\begin{equation}
\beta\crit = 0.64805(41)\ \Rightarrow\  T\crit = 1.5431(10)\ ,
\end{equation}
with $\chi^2/\text{dof} = 4.47/10$. Note that the contribution to the error
from the uncertainty in $\omega$ is very small.

\subsection{One-dimensional long range model with $\boldsymbol{\sigma =0.896}$}
\label{sec:res-896}

\begin{figure}
\includegraphics[width=\columnwidth]{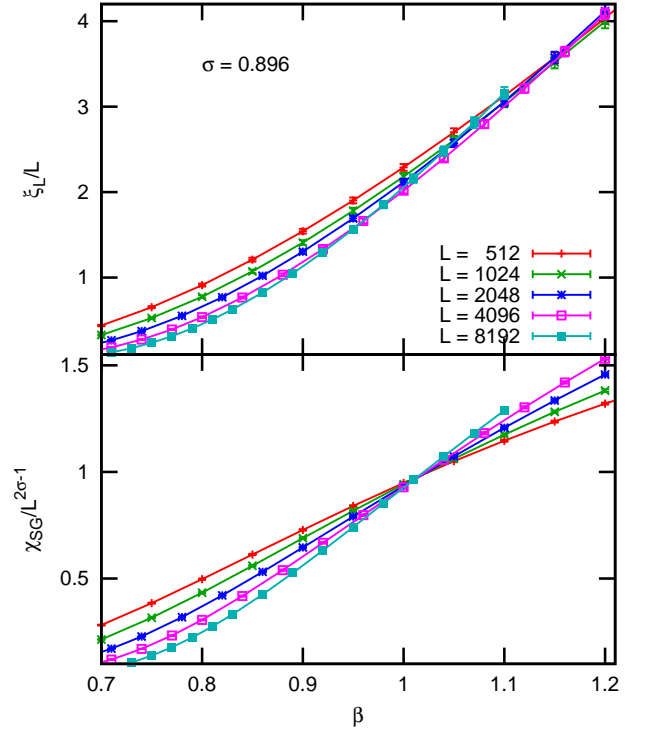}
\caption{(Color online) Correlation length in units of the system size ({\bf
top}) and scale-invariant combination of the SG susceptibility and the
lattice dimension $\chi_\text{SG}/L^{2\sigma-1}$ ({\bf bottom}), as a function of the
inverse temperature $\beta$, for the LR-model with
$\sigma=0.896$. For both quantities, the curves
for the different $L$ should cross at temperatures that approach the
critical point when $L$ grows, see Eq.~\eqref{t*}.  }
\label{fig:scaleinvariant_896}
\end{figure}

According to Eq.~\eqref{sigma_eta} and the value of $\eta$ for the 3-$d$ model
given in Ref.~\onlinecite{hasenbusch:08b}, $\eta_\mathrm{SR}(3) = -0.375(10)$,
$\sigma = 0.896$
is a proxy for 3-$d$, at least according to the comparison of the
exponents $\eta$ (or equivalently of the magnetic exponents $y_H$). We now
attempt to see if the correspondence also works for the exponents $\omega$ and
$\nu$.

As we show in Fig.~\ref{fig:scaleinvariant_896}, $\xi_L/L$ displays a rather marginal
behavior for this value of $\sigma$. We are not able to resolve the
crossing temperatures for this dimensionless quantity. On the other hand,
crossing points of $\chi_\text{SG}/L^{2\sigma-1}$ are easily identified. Our
interpretation of these findings is that, for this value of $\sigma$, we are
fairly close to the critical value $\sigma_l$,
such that for $\sigma>\sigma_l$ there is no longer a SG
phase, see Sec.~\ref{sec:intro}. It is expected that\cite{kotliar:83}
$\sigma_l = 1$ since this corresponds to $d -2 + \eta = 0$ with $d =1$ and
$\eta = \eta_{LR}(\sigma) = 3 - 2\sigma$. Hence a transition is expected for
$\sigma = 0.896$. It is easier to find crossing points from
$\chi_\text{SG}/L^{2\sigma-1}$, because, in the SG phase, it scales as $L^a$
with an exponent $a$ larger than the corresponding one for $\xi_L/L$, so we
feel that our results for $\sigma = 0.896$ are consistent with the expected
transition.

\begin{figure}
\includegraphics[width=\columnwidth]{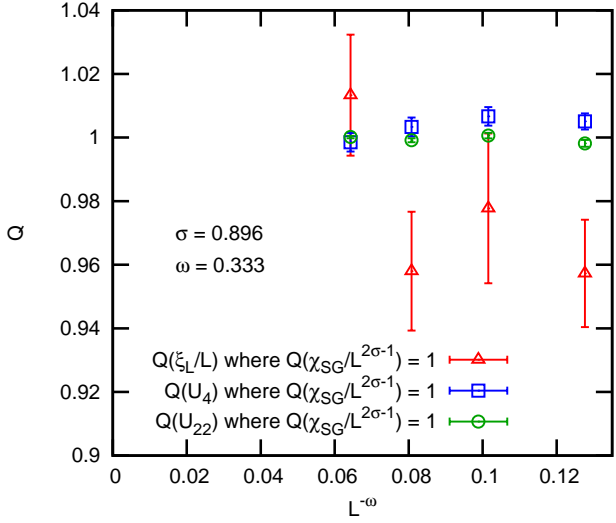}
\caption{(Color online)
Quotients of the dimensionless quantities $\xi_L/L, U_4$ and $U_{22}$ at the
crossings of $\chi_{SG}/L^{2\sigma-1}$ for $\sigma = 0.896$.
Compared to the error bars there is
very little size dependence so the data is inadequate to determine the
correction to scaling exponent $\omega$.
}
\label{fig:Q_omega_896}
\end{figure}

Unfortunately, plots of dimensionless quantities do not allow us to determine $\omega$
because there is very little size dependence in the quotients. This is
illustrated in Fig.~\ref{fig:Q_omega_896} which shows quotients of $\xi_L/L,
U_4$ and $U_{22}$ at crossings of $\chi_{SG}/L^{2\sigma - 1}$.

\begin{figure}
\includegraphics[width=\columnwidth]{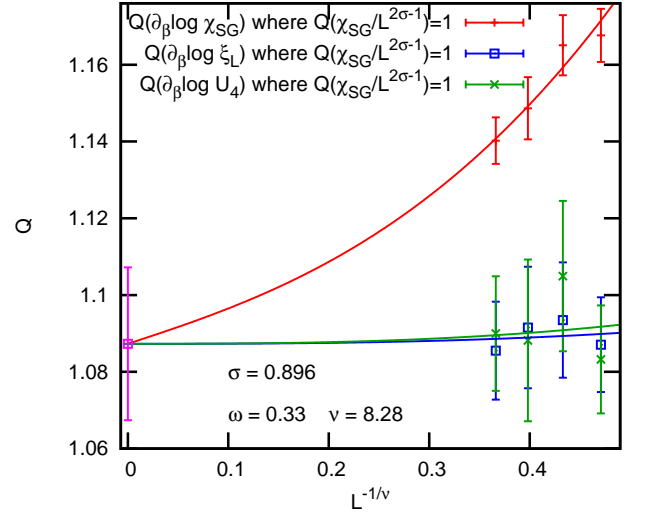}
\caption{(Color online) The quotients of $\partial_\beta \log \xi_L$,
$\partial_\beta \log U_4$ and $\partial_\beta \log \chi_{SG}$ at the
crossings of $\chi_{SG}/L^{2\sigma-1}$ for $\sigma = 0.896$.  For $\partial_\beta \log \xi_L$ and 
$\partial_\beta \log U_4$ the lines are fits to functions of type $Q+B_1 L^{-\omega}$, where
$Q = 2^{1/\nu}$, see Eq.~\eqref{QUOTIENTS}.
For $\chi_{SG}$ we need to consider
also an $L^{-1/\nu}$ term, see Eq.~\eqref{quotient_chi}.
The $\omega$ value is fixed to the value 0.33 
expected from the 3-$d$ data of Hasenbusch et
al.\cite{hasenbusch:08b} who find $\omega_{SR}(3) = 1.0(1)$,
and Eq.~\eqref{omega_compare}, while the
value of $\nu$ is a fit parameter.
}
\label{fig:Q_nu_fix_w_896}
\end{figure}

To determine $\nu$ we first consider the quotients of the logarithmic
derivatives with respect to $\beta$ of the dimensionless quantities $\xi_L$ and
$U_4$ at the $\chi_\text{SG}/L^{2\sigma-1}$ crossing. A fit to $Q+B_1
L^{-\omega}$, does not allow us to find $\omega$, so we fix
the value $\omega=0.33(3)$, obtained from Eq.~\eqref{SR_LR} and the result of
Hasenbusch et al.\cite{hasenbusch:08b} that $\omega_{SR}(3) = 1.0(1)$,
obtaining
\begin{equation}
Q \equiv 2^{1/\nu} = 1.0890(202)[2]\,,\ \chi^2/\text{dof}=1.14/5\ ,
\end{equation}
which determines $\nu$ to be in the range $5.7<\nu<10.4$. Notice the smallness
of the error bars coming from the $\omega$ error, or conversely the difficulty
of determining $\omega$ from these quantities.

We also tried a more complex fit including the quotients of logarithmic
derivatives of the scale invariant quantity $\chi_\text{SG}/L^{2\sigma-1}$. 
As discussed in
Sec.~\ref{sec:fss}, this derivative (but only this one) suffers from
additional scaling corrections of order
$L^{-1/\nu}$. Note that, according to Eqs.~(9) and (10) and the
SR values\cite{hasenbusch:08b} $\omega_{SR}(3) = 1.0(1), \nu_{SR}(3) = 
2.45(15)$, we expect
$\omega_{LR}\approx 0.33$ and $1/\nu_{LR}\approx 0.14$, so the corrections of
order $L^{-1/\nu}$ are dominant. We therefore fit the data for the quotients
of the logarithmic
derivative of  $\chi_\text{SG}/L^{2\sigma-1}$ to Eq.~\eqref{quotient_chi},
while for the quotients of the logarithmic derivatives of 
$U_4$ and $\xi_L/L$ we use Eq.~\eqref{QUOTIENTS} with $y_O = 1/\nu$, which corresponds to $B_2=0$.

To obtain a reliable fit, we need fix the value of $\omega$ and, as above, we
take this to be
$\omega=0.33(3)$, obtaining
\begin{equation}
Q \equiv 2^{1/\nu} =1.087(199)[3],\ \chi^2/\text{dof}=1.54/7\ ,
\end{equation}
which determines $\nu$ to be in the range $6.8<\nu<10.6$, so our estimate for
$\nu$ is
\begin{equation}
\nu_{LR}(0.896) = 8.7(1.9). 
\end{equation}
Again, the effect of
the $\omega$ uncertainty is very small. We have tried to bound the $\omega$
value from this fit, but the result is almost useless [$\omega\in(0,0.97)$].

Finally we discuss the value of $\beta\crit$. We do not see any evolution of
$\beta\crit$ with $L$. However we perform several fits
to estimate the extrapolation errors.
First we try a fit of the $\chi_{SG}/L^{2\sigma-1}$ crossings taking
$\omega$ and $\nu$
from
the 3-$d$ derived values: $\omega=0.33(3)$, $\nu=7.35(45)$ so
$\omega+1/\nu=0.47(4)$. The result is
$\beta\crit=1.004(15)[1]$, with $\chi^2/\text{dof}=0.24/2$.
If we use $\omega=0.33(3)$ but the $\nu$ value obtained above
$\nu=8.7(1.9)$, i.e.\ $\omega+1/\nu=0.44(5)$ we get
$ \beta\crit=1.003(16)[2], \chi^2/\text{dof}=0.23/2$.

These last two results are statistically correlated, and
we take the latter as our final estimate:
\begin{equation}
\beta\crit = 1.003(18) \ \Rightarrow \ T\crit = 0.997(18) \, .
\end{equation}

\section{Conclusions}
\label{sec:conclusions}

The purpose of this paper is to see if there is a value of $\sigma$ for the
long-range spin glass model which corresponds precisely to a short-range
four-dimensional spin glass, and (with a different value of $\sigma$) to a
three-dimensional spin glass, in the sense that \textit{all} the LR and SR exponents, in
particular, $\eta, \nu$ and $\omega$, match in the sense of
Eqs.~\eqref{SR_LR}--\eqref{omega_compare}. Since $\eta_\mathrm{LR}$ is given exactly by
the simple expression in Eq.~\eqref{eta_LR}, we have chosen two values of
$\sigma, 0.790$ and $0.896$, as proxies for 4-$d$ and 3-$d$
respectively, since the values of $\eta$ match according to
Eq.~\eqref{sigma_eta}. The question, then, is whether the \textit{other}
exponents, $\omega$ and $\nu$, match according to Eqs.~\eqref{omega_compare}
and \eqref{nu_compare}.

Our results for $\omega$ and $\nu$ are summarized in Table
\ref{tab:exponents}. For the case of 4-$d$, the correspondence works well, the
values for the exponents being consistent with
Eqs.~\eqref{nu_compare}
and \eqref{omega_compare} within reasonably modest error bars. However, for 3-$d$, we
are not able to establish a sharp connection, since, for the corresponding
long-range model, $\sigma = 0.896$, we can not determine $\omega$. If
we \textit{assume} that the value of $\omega_\mathrm{LR}(0.896)$ is that given by the matching formula,
Eq.~\eqref{omega_compare}, with the value of $\omega$ from the 3-$d$
simulations,\cite{hasenbusch:08b} namely $\omega_\mathrm{LR}(0.896) = 0.33(3)$, then we find
$\nu_\text{LR}= 8.7 \pm 1.9$ which \textit{is} consistent with
$3 \nu_\text{SR}(3) = 7.35 \pm  0.45$.

While it seems unlikely to us that all the critical exponents of the LR and SR
models match \textit{exactly}
according to Eq.~\eqref{SR_LR}, our results indicate
that these equations are satisfied to a good approximation, and hence the
critical behavior of the SR and corresponding LR models are very similar.
Whether this similarity extends to the more subtle question of the nature of
the spin glass phase below $T\crit$ remains to be seen.

\begin{table}
\caption{
Summary of results for critical exponents of the short-range models in 3-$d$
and 4-$d$, the expected (proxy) results for the long-range models based on the
short-range results and the connection in Eq.~\eqref{SR_LR}, and the actual
results for the long-range models. It was not possible to estimate $\omega$
for the long-range model with $\sigma =0.896$. If we assume that it is given by the
matching formula, $\omega_\mathrm{SR}/d$, then we obtain the result for
$\nu_\mathrm{LR}(0.896)$ shown in the
table. The 3-$d$ results are from
Ref.~\onlinecite{hasenbusch:08b}, and all other results are from the present work.
}
\label{tab:exponents}
\begin{tabular*}{\columnwidth}{@{\extracolsep{\fill}}ldd}
\hline
\hline
&\multicolumn{1}{r}{$d=4,\sigma=0.790$} &\multicolumn{1}{r}{$d=3,\sigma=0.896$}\\
\hline
$\omega_\mathrm{SR}(d)$     & 1.04(10)           &     1.0(1)   \\
$\omega_\mathrm{SR}(d) / d$ & 0.26(4)            &     0.33(3)  \\
$\omega_\mathrm{LR}(\sigma)$& 0.277(8)           & \multicolumn{1}{c}{---}     \\
\hline
$\nu_\mathrm{SR}(d)$        & 1.068(5)           &     2.45(15) \\
$d\, \nu_\mathrm{SR}(d)$    & 4.272(20)          &     7.35(45) \\
$\nu_\mathrm{LR}(\sigma)$   & 4.41(19)           &     8.7(1.9)\\
\hline
\hline
\end{tabular*}
\end{table}

\begin{acknowledgments}
We thank G. Parisi and M. Moore for discussions.  APY acknowledges support
from the NSF through grant No.~DMR-0906366 and a generous allocation of
computer time from the Hierarchical Systems Research Foundation. The short range
simulations, and part of long range simulations, have been carried out in
ARAGRID and BIFI computers. RAB, LAF and
VMM acknowledge partial financial support from MICINN, Spain, contract
FIS2009-12648-C03. RAB was also supported by the FPI program (Diputaci\'on de
Arag\'on, Spain). VMM thanks the hospitality of the Physics Department of UCSC
(visit funded by the \emph{del Amo} foundation), where part of this work was
performed.

\end{acknowledgments}

\bibliography{refs,comments}

\begin{thebibliography}{31}
\expandafter\ifx\csname natexlab\endcsname\relax\def\natexlab#1{#1}\fi
\expandafter\ifx\csname bibnamefont\endcsname\relax
  \def\bibnamefont#1{#1}\fi
\expandafter\ifx\csname bibfnamefont\endcsname\relax
  \def\bibfnamefont#1{#1}\fi
\expandafter\ifx\csname citenamefont\endcsname\relax
  \def\citenamefont#1{#1}\fi
\expandafter\ifx\csname url\endcsname\relax
  \def\url#1{\texttt{#1}}\fi
\expandafter\ifx\csname urlprefix\endcsname\relax\def\urlprefix{URL }\fi
\providecommand{\bibinfo}[2]{#2}
\providecommand{\eprint}[2][]{\url{#2}}

\bibitem[{\citenamefont{Binder and Young}(1986)}]{binder:86}
\bibinfo{author}{\bibfnamefont{K.}~\bibnamefont{Binder}} \bibnamefont{and}
  \bibinfo{author}{\bibfnamefont{A.~P.} \bibnamefont{Young}},
  \emph{\bibinfo{title}{Spin glasses: Experimental facts, theoretical concepts
  and open questions}}, \bibinfo{journal}{Rev. Mod. Phys.}
  \textbf{\bibinfo{volume}{58}}, \bibinfo{pages}{801} (\bibinfo{year}{1986}).

\bibitem[{\citenamefont{Privman}(1990)}]{privman:90}
\bibinfo{editor}{\bibfnamefont{V.}~\bibnamefont{Privman}}, ed.,
  \emph{\bibinfo{title}{Finite Size Scaling and Numerical Simulation of
  Statistical Systems}} (\bibinfo{publisher}{World Scientific},
  \bibinfo{address}{Singapore}, \bibinfo{year}{1990}).

\bibitem[{\citenamefont{Amit and Matin-Mayor}(2005)}]{amit:05}
\bibinfo{author}{\bibfnamefont{D.}~\bibnamefont{Amit}} \bibnamefont{and}
  \bibinfo{author}{\bibfnamefont{V.}~\bibnamefont{Matin-Mayor}},
  \emph{\bibinfo{title}{Field Theory, the Renormalization Group and Critical
  Phenomena}} (\bibinfo{publisher}{World Scientific},
  \bibinfo{address}{Singapore}, \bibinfo{year}{2005}).

\bibitem[{\citenamefont{Katzgraber and
  Young}(2003{\natexlab{a}})}]{katzgraber:03}
\bibinfo{author}{\bibfnamefont{H.~G.} \bibnamefont{Katzgraber}}
  \bibnamefont{and} \bibinfo{author}{\bibfnamefont{A.~P.} \bibnamefont{Young}},
  \emph{\bibinfo{title}{Monte {C}arlo studies of the one-dimensional {I}sing
  spin glass with power-law interactions}}, \bibinfo{journal}{Phys, Rev. B}
  \textbf{\bibinfo{volume}{67}}, \bibinfo{pages}{134410}
  (\bibinfo{year}{2003}{\natexlab{a}}).

\bibitem[{\citenamefont{Katzgraber and
  Young}(2003{\natexlab{b}})}]{katzgraber:03b}
\bibinfo{author}{\bibfnamefont{H.~G.} \bibnamefont{Katzgraber}}
  \bibnamefont{and} \bibinfo{author}{\bibfnamefont{A.~P.} \bibnamefont{Young}},
  \emph{\bibinfo{title}{Geometry of large-scale low-energy excitations in the
  one-dimensional {I}sing spin glass with power-law interactions}},
  \bibinfo{journal}{Phys, Rev. B} \textbf{\bibinfo{volume}{68}},
  \bibinfo{pages}{224408} (\bibinfo{year}{2003}{\natexlab{b}}),
  \eprint{(arXiv:cond-mat/0307583)}.

\bibitem[{\citenamefont{Katzgraber and Young}(2005)}]{katzgraber:05}
\bibinfo{author}{\bibfnamefont{H.~G.} \bibnamefont{Katzgraber}}
  \bibnamefont{and} \bibinfo{author}{\bibfnamefont{A.~P.} \bibnamefont{Young}},
  \emph{\bibinfo{title}{Probing the {A}lmeida-{T}houless line away from the
  mean-field model}}, \bibinfo{journal}{Phys. Rev. B}
  \textbf{\bibinfo{volume}{72}}, \bibinfo{pages}{184416}
  (\bibinfo{year}{2005}).

\bibitem[{\citenamefont{Katzgraber et~al.}(2009)\citenamefont{Katzgraber,
  Larson, and Young}}]{katzgraber:09}
\bibinfo{author}{\bibfnamefont{H.~G.} \bibnamefont{Katzgraber}},
  \bibinfo{author}{\bibfnamefont{D.}~\bibnamefont{Larson}}, \bibnamefont{and}
  \bibinfo{author}{\bibfnamefont{A.~P.} \bibnamefont{Young}},
  \emph{\bibinfo{title}{Study of the de {A}lmeida-{T}houless line using
  power-law diluted one-dimensional {I}sing spin glasses}},
  \bibinfo{journal}{Phys. Rev. Lett} \textbf{\bibinfo{volume}{102}},
  \bibinfo{pages}{177205} (\bibinfo{year}{2009}), \eprint{(arXiv:0812:0421)}.

\bibitem[{\citenamefont{Larson et~al.}(2010)\citenamefont{Larson, Katzgraber,
  Moore, and Young}}]{larson:10}
\bibinfo{author}{\bibfnamefont{D.}~\bibnamefont{Larson}},
  \bibinfo{author}{\bibfnamefont{H.~G.} \bibnamefont{Katzgraber}},
  \bibinfo{author}{\bibfnamefont{M.~A.} \bibnamefont{Moore}}, \bibnamefont{and}
  \bibinfo{author}{\bibfnamefont{A.~P.} \bibnamefont{Young}},
  \emph{\bibinfo{title}{Numerical studies of a one-dimensional 3-spin
  spin-glass model with long-range interactions}}, \bibinfo{journal}{Phys. Rev.
  B} \textbf{\bibinfo{volume}{81}}, \bibinfo{pages}{064415}
  (\bibinfo{year}{2010}), \eprint{(arXiv:0908.2224)}.

\bibitem[{\citenamefont{Sharma and Young}(2011)}]{sharma:11a}
\bibinfo{author}{\bibfnamefont{A.}~\bibnamefont{Sharma}} \bibnamefont{and}
  \bibinfo{author}{\bibfnamefont{A.~P.} \bibnamefont{Young}},
  \emph{\bibinfo{title}{Phase {T}ransitions in the 1-d {L}ong-{R}ange {D}iluted
  {H}eisenberg {S}pin {G}lass}} (\bibinfo{year}{2011}),
  \eprint{(arXiv:1103.3297)}.

\bibitem[{\citenamefont{Leuzzi et~al.}(2008)\citenamefont{Leuzzi, Parisi,
  Ricci-Tersenghi, and Ruiz-Lorenzo}}]{leuzzi:08}
\bibinfo{author}{\bibfnamefont{L.}~\bibnamefont{Leuzzi}},
  \bibinfo{author}{\bibfnamefont{G.}~\bibnamefont{Parisi}},
  \bibinfo{author}{\bibfnamefont{F.}~\bibnamefont{Ricci-Tersenghi}},
  \bibnamefont{and} \bibinfo{author}{\bibfnamefont{J.~J.}
  \bibnamefont{Ruiz-Lorenzo}}, \emph{\bibinfo{title}{Diluted one-dimensional
  spin glasses with power law decaying interactions}}, \bibinfo{journal}{Phys.
  Rev. Lett} \textbf{\bibinfo{volume}{101}}, \bibinfo{pages}{107203}
  (\bibinfo{year}{2008}).

\bibitem[{\citenamefont{Leuzzi et~al.}(2009)\citenamefont{Leuzzi, Parisi,
  Ricci-Tersenghi, and Ruiz-Lorenzo}}]{leuzzi:09}
\bibinfo{author}{\bibfnamefont{L.}~\bibnamefont{Leuzzi}},
  \bibinfo{author}{\bibfnamefont{G.}~\bibnamefont{Parisi}},
  \bibinfo{author}{\bibfnamefont{F.}~\bibnamefont{Ricci-Tersenghi}},
  \bibnamefont{and} \bibinfo{author}{\bibfnamefont{J.~J.}
  \bibnamefont{Ruiz-Lorenzo}}, \emph{\bibinfo{title}{Ising spin-glass
  transition in a magnetic field outside the limit of validity of mean-field
  theory}}, \bibinfo{journal}{Phys. Rev. Lett} \textbf{\bibinfo{volume}{103}},
  \bibinfo{pages}{267201} (\bibinfo{year}{2009}).

\bibitem[{\citenamefont{Leuzzi et~al.}(2011)\citenamefont{Leuzzi, Parisi,
  Ricci-Tersenghi, and Ruiz-Lorenzo}}]{leuzzi:11}
\bibinfo{author}{\bibfnamefont{L.}~\bibnamefont{Leuzzi}},
  \bibinfo{author}{\bibfnamefont{G.}~\bibnamefont{Parisi}},
  \bibinfo{author}{\bibfnamefont{F.}~\bibnamefont{Ricci-Tersenghi}},
  \bibnamefont{and} \bibinfo{author}{\bibfnamefont{J.~J.}
  \bibnamefont{Ruiz-Lorenzo}}, \emph{\bibinfo{title}{Bond diluted levy
  spin-glass model and a new finite size scaling method to determine a phase
  transition}}, \bibinfo{journal}{Philos. Mag.} \textbf{\bibinfo{volume}{91}},
  \bibinfo{pages}{1917} (\bibinfo{year}{2011}).

\bibitem[{\citenamefont{Harris et~al.}(1976)\citenamefont{Harris, Lubensky, and
  Chen}}]{harris:76}
\bibinfo{author}{\bibfnamefont{A.~B.} \bibnamefont{Harris}},
  \bibinfo{author}{\bibfnamefont{T.~C.} \bibnamefont{Lubensky}},
  \bibnamefont{and} \bibinfo{author}{\bibfnamefont{J.-H.} \bibnamefont{Chen}},
  \emph{\bibinfo{title}{Critical properties of spin-glasses}},
  \bibinfo{journal}{Phys. Rev. Lett.} \textbf{\bibinfo{volume}{36}},
  \bibinfo{pages}{415} (\bibinfo{year}{1976}).

\bibitem[{\citenamefont{{Kotliar} et~al.}(1983)\citenamefont{{Kotliar},
  {Anderson}, and {Stein}}}]{kotliar:83}
\bibinfo{author}{\bibfnamefont{G.}~\bibnamefont{{Kotliar}}},
  \bibinfo{author}{\bibfnamefont{P.~W.} \bibnamefont{{Anderson}}},
  \bibnamefont{and} \bibinfo{author}{\bibfnamefont{D.~L.}
  \bibnamefont{{Stein}}}, \emph{\bibinfo{title}{One-dimensional spin-glass
  model with long-range random interactions}}, \bibinfo{journal}{Phys. Rev. B}
  \textbf{\bibinfo{volume}{27}}, \bibinfo{pages}{602} (\bibinfo{year}{1983}).

\bibitem[{moo()}]{moore:pc}
\bibinfo{note}{M.A.~Moore (private communication)}.

\bibitem[{par()}]{parisi:pc}
\bibinfo{note}{G.~Parisi (private communication)}.

\bibitem[{\citenamefont{Fisher et~al.}(1972)\citenamefont{Fisher, Ma, and
  Nickel}}]{fisher:72}
\bibinfo{author}{\bibfnamefont{M.~E.} \bibnamefont{Fisher}},
  \bibinfo{author}{\bibfnamefont{S.-k.} \bibnamefont{Ma}}, \bibnamefont{and}
  \bibinfo{author}{\bibfnamefont{B.~G.} \bibnamefont{Nickel}},
  \emph{\bibinfo{title}{Critical exponents for long-range interactions}},
  \bibinfo{journal}{Phys. Rev. Lett.} \textbf{\bibinfo{volume}{29}},
  \bibinfo{pages}{917} (\bibinfo{year}{1972}).

\bibitem[{\citenamefont{Hasenbusch et~al.}(2008)\citenamefont{Hasenbusch,
  Pelissetto, and Vicari}}]{hasenbusch:08b}
\bibinfo{author}{\bibfnamefont{M.}~\bibnamefont{Hasenbusch}},
  \bibinfo{author}{\bibfnamefont{A.}~\bibnamefont{Pelissetto}},
  \bibnamefont{and} \bibinfo{author}{\bibfnamefont{E.}~\bibnamefont{Vicari}},
  \emph{\bibinfo{title}{The critical behavior of three-dimensional {I}sing
  glass models}}, \bibinfo{journal}{Phys. Rev. B}
  \textbf{\bibinfo{volume}{78}}, \bibinfo{pages}{214205}
  (\bibinfo{year}{2008}), \eprint{(arXiv:0809.3329)}.

\bibitem[{\citenamefont{Newman and Barkema}(1999)}]{newman:99}
\bibinfo{author}{\bibfnamefont{M.~E.~J.} \bibnamefont{Newman}}
  \bibnamefont{and} \bibinfo{author}{\bibfnamefont{G.~T.}
  \bibnamefont{Barkema}}, \emph{\bibinfo{title}{{M}onte {C}arlo Methods in
  Statistical Physics}} (\bibinfo{publisher}{Oxford University Press Inc.},
  \bibinfo{address}{New York, USA}, \bibinfo{year}{1999}).

\bibitem[{\citenamefont{Fern\'andez et~al.}(2010)\citenamefont{Fern\'andez,
  Mart\'{\i}n-Mayor, Parisi, and Seoane}}]{fernandez:10}
\bibinfo{author}{\bibfnamefont{L.~A.} \bibnamefont{Fern\'andez}},
  \bibinfo{author}{\bibfnamefont{V.}~\bibnamefont{Mart\'{\i}n-Mayor}},
  \bibinfo{author}{\bibfnamefont{G.}~\bibnamefont{Parisi}}, \bibnamefont{and}
  \bibinfo{author}{\bibfnamefont{B.}~\bibnamefont{Seoane}},
  \emph{\bibinfo{title}{Spin glasses on the hypercube}},
  \bibinfo{journal}{Phys. Rev. B} \textbf{\bibinfo{volume}{81}},
  \bibinfo{pages}{134403} (\bibinfo{year}{2010}).

\bibitem[{\citenamefont{Cooper et~al.}(1982)\citenamefont{Cooper, Freedman, and
  Preston}}]{cooper:82}
\bibinfo{author}{\bibfnamefont{B.}~\bibnamefont{Cooper}},
  \bibinfo{author}{\bibfnamefont{B.}~\bibnamefont{Freedman}}, \bibnamefont{and}
  \bibinfo{author}{\bibfnamefont{D.}~\bibnamefont{Preston}},
  \emph{\bibinfo{title}{Solving $\varphi^4_{1,2}$ field theory with {M}onte
  {C}arlo}}, \bibinfo{journal}{Nucl. Phys. B.} \textbf{\bibinfo{volume}{210}},
  \bibinfo{pages}{210} (\bibinfo{year}{1982}).

\bibitem[{\citenamefont{Palassini and Caracciolo}(1999)}]{palassini:99b}
\bibinfo{author}{\bibfnamefont{M.}~\bibnamefont{Palassini}} \bibnamefont{and}
  \bibinfo{author}{\bibfnamefont{S.}~\bibnamefont{Caracciolo}},
  \emph{\bibinfo{title}{Universal finite size scaling functions in the 3d
  {I}sing spin glass}}, \bibinfo{journal}{Phys. Rev. Lett.}
  \textbf{\bibinfo{volume}{82}}, \bibinfo{pages}{5128} (\bibinfo{year}{1999}),
  \eprint{(arXiv:cond-mat/9904246)}.

\bibitem[{\citenamefont{Ballesteros et~al.}(2000)\citenamefont{Ballesteros,
  Cruz, Fernandez, Martin-Mayor, Pech, Ruiz-Lorenzo, Tarancon, Tellez, Ullod,
  and Ungil}}]{ballesteros:00}
\bibinfo{author}{\bibfnamefont{H.~G.} \bibnamefont{Ballesteros}},
  \bibinfo{author}{\bibfnamefont{A.}~\bibnamefont{Cruz}},
  \bibinfo{author}{\bibfnamefont{L.~A.} \bibnamefont{Fernandez}},
  \bibinfo{author}{\bibfnamefont{V.}~\bibnamefont{Martin-Mayor}},
  \bibinfo{author}{\bibfnamefont{J.}~\bibnamefont{Pech}},
  \bibinfo{author}{\bibfnamefont{J.~J.} \bibnamefont{Ruiz-Lorenzo}},
  \bibinfo{author}{\bibfnamefont{A.}~\bibnamefont{Tarancon}},
  \bibinfo{author}{\bibfnamefont{P.}~\bibnamefont{Tellez}},
  \bibinfo{author}{\bibfnamefont{C.~L.} \bibnamefont{Ullod}}, \bibnamefont{and}
  \bibinfo{author}{\bibfnamefont{C.}~\bibnamefont{Ungil}},
  \emph{\bibinfo{title}{Critical behavior of the three-dimensional {I}sing spin
  glass}}, \bibinfo{journal}{Phys. Rev. B} \textbf{\bibinfo{volume}{62}},
  \bibinfo{pages}{14237} (\bibinfo{year}{2000}),
  \eprint{(arXiv:cond-mat/0006211)}.

\bibitem[{\citenamefont{Binder}(1981)}]{binder:81b}
\bibinfo{author}{\bibfnamefont{K.}~\bibnamefont{Binder}},
  \emph{\bibinfo{title}{Finite size scaling analysis of {I}sing model block
  distribution functions}}, \bibinfo{journal}{Z. Phys. B}
  \textbf{\bibinfo{volume}{43}}, \bibinfo{pages}{119} (\bibinfo{year}{1981}).

\bibitem[{\citenamefont{Ballesteros et~al.}(1996)\citenamefont{Ballesteros,
  Fernandez, Martin-Mayor, Pech, and Mu\~noz Sudupe}}]{ballesteros:96a}
\bibinfo{author}{\bibfnamefont{H.~G.} \bibnamefont{Ballesteros}},
  \bibinfo{author}{\bibfnamefont{L.~A.} \bibnamefont{Fernandez}},
  \bibinfo{author}{\bibfnamefont{V.}~\bibnamefont{Martin-Mayor}},
  \bibinfo{author}{\bibfnamefont{J.}~\bibnamefont{Pech}}, \bibnamefont{and}
  \bibinfo{author}{\bibfnamefont{A.}~\bibnamefont{Mu\~noz Sudupe}},
  \emph{\bibinfo{title}{Finite size effects on measures of critical exponents
  in d=3 {O(N)} models}}, \bibinfo{journal}{Phys. Lett. B}
  \textbf{\bibinfo{volume}{387}}, \bibinfo{pages}{125} (\bibinfo{year}{1996}),
  \eprint{(arXiv:cond-mat/9606203)}.

\bibitem[{\citenamefont{Nightingale}(1976)}]{nightingale:76}
\bibinfo{author}{\bibfnamefont{M.~P.} \bibnamefont{Nightingale}},
  \emph{\bibinfo{title}{Scaling theory and finite systems}},
  \bibinfo{journal}{Physica A} \textbf{\bibinfo{volume}{83}},
  \bibinfo{pages}{561} (\bibinfo{year}{1976}).

\bibitem[{\citenamefont{Ballesteros et~al.}(1998)\citenamefont{Ballesteros,
  Fern\'andez, Martin-Mayor, Mu\~noz Sudupe, Parisi, and
  Ruiz-Lorenzo}}]{ballesteros:98}
\bibinfo{author}{\bibfnamefont{H.~G.} \bibnamefont{Ballesteros}},
  \bibinfo{author}{\bibfnamefont{L.~A.} \bibnamefont{Fern\'andez}},
  \bibinfo{author}{\bibfnamefont{V.}~\bibnamefont{Martin-Mayor}},
  \bibinfo{author}{\bibfnamefont{A.}~\bibnamefont{Mu\~noz Sudupe}},
  \bibinfo{author}{\bibfnamefont{G.}~\bibnamefont{Parisi}}, \bibnamefont{and}
  \bibinfo{author}{\bibfnamefont{J.~J.} \bibnamefont{Ruiz-Lorenzo}},
  \emph{\bibinfo{title}{Critical exponents of the three-dimensional diluted
  {I}sing model}}, \bibinfo{journal}{Phys. Rev. B}
  \textbf{\bibinfo{volume}{58}}, \bibinfo{pages}{2740} (\bibinfo{year}{1998}).

\bibitem[{\citenamefont{Weigel and Janke}(2009)}]{weigel:09}
\bibinfo{author}{\bibfnamefont{M.}~\bibnamefont{Weigel}} \bibnamefont{and}
  \bibinfo{author}{\bibfnamefont{W.}~\bibnamefont{Janke}},
  \emph{\bibinfo{title}{Cross correlations in scaling analyses of phase
  transitions}}, \bibinfo{journal}{Phys. Rev. Lett.}
  \textbf{\bibinfo{volume}{102}}, \bibinfo{pages}{100601}
  (\bibinfo{year}{2009}).

\bibitem[{\citenamefont{Hukushima and Nemoto}(1996)}]{hukushima:96}
\bibinfo{author}{\bibfnamefont{K.}~\bibnamefont{Hukushima}} \bibnamefont{and}
  \bibinfo{author}{\bibfnamefont{K.}~\bibnamefont{Nemoto}},
  \emph{\bibinfo{title}{Exchange {M}onte {C}arlo method and application to spin
  glass simulations}}, \bibinfo{journal}{J. Phys. Soc. Japan}
  \textbf{\bibinfo{volume}{65}}, \bibinfo{pages}{1604} (\bibinfo{year}{1996}),
  \eprint{(arXiv:cond-mat/9512035)}.

\bibitem[{\citenamefont{Marinari and Zuliani}(1999)}]{marinari:99}
\bibinfo{author}{\bibfnamefont{E.}~\bibnamefont{Marinari}} \bibnamefont{and}
  \bibinfo{author}{\bibfnamefont{F.}~\bibnamefont{Zuliani}},
  \emph{\bibinfo{title}{Numerical simulations of the 4d {E}dwards-{A}nderson
  spin glass with binary couplings}}, \bibinfo{journal}{J. Phys. A}
  \textbf{\bibinfo{volume}{32}}, \bibinfo{pages}{7447} (\bibinfo{year}{1999}).

\bibitem[{\citenamefont{J\"org and Katzgraber}(2008)}]{jorg:08c}
\bibinfo{author}{\bibfnamefont{T.}~\bibnamefont{J\"org}} \bibnamefont{and}
  \bibinfo{author}{\bibfnamefont{H.~G.} \bibnamefont{Katzgraber}},
  \emph{\bibinfo{title}{Universality and universal finite-size scaling
  functions in four-dimensional ising spin glasses numerical simulations of the
  4d}}, \bibinfo{journal}{Phys. Rev. B} \textbf{\bibinfo{volume}{77}},
  \bibinfo{pages}{214426} (\bibinfo{year}{2008}).

\end{thebibliography}

\end{document}